\documentclass[12pt]{article}
\usepackage{graphicx} 
\usepackage{amsmath}
\usepackage{amssymb}
\usepackage{csquotes}
\usepackage{natbib}
\usepackage{longtable}
\usepackage{float}
\usepackage{xcolor}
\usepackage{algorithm}
\usepackage{algorithmic}
\usepackage{bm}
\usepackage{actuarialsymbol}
\usepackage{multirow}
\usepackage{caption}    
\usepackage{booktabs}
\usepackage{subcaption}

\newtheorem{definition}{Definition}

\newcommand{\Ab}{\mathbf{A}}

\newcommand{\Cb}{\mathbf{C}}
\newcommand{\Eb}{\mathbf{E}}
\newcommand{\Fb}{\mathbf{F}}
\newcommand{\Hb}{\mathbf{H}}
\newcommand{\Ib}{\mathbf{I}}

\newcommand{\Wb}{\mathbf{W}}

\newcommand{\Yb}{\mathbf{Y}}
\newcommand{\RR}{\mathbb{R}}
\newcommand{\cL}{\mathcal{L}}
\newcommand{\cG}{\mathcal{G}}
\newcommand{\cP}{\mathcal{P}}

\newcommand{\ba}{\bm{a}}
\newcommand{\bx}{\bm{x}}
\newcommand{\by}{\bm{y}}

\newcommand{\bbm}{\bm{m}}
\newcommand{\bbf}{\bm{f}}

\newcommand{\bepsilon}{\bm{\epsilon}}
\newcommand{\blambda}{\bm{\lambda}}
\newcommand{\bLambda}{\bm{\Lambda}}

\newcommand{\argmin}{\mathop{\mathrm{argmin}}}

\makeatletter
\renewcommand{\paragraph}{\@startsection{paragraph}{4}{0ex}%
   {-3.25ex plus -1ex minus -0.2ex}%
   {1.5ex plus 0.2ex}%
   {\normalfont\normalsize\bfseries}}
\makeatother
\stepcounter{secnumdepth}
\stepcounter{tocdepth}
\renewcommand{\arraystretch}{1.5} 

\usepackage{setspace}
\usepackage[margin=1in]{geometry}
\usepackage{authblk}

\begin{document}

\title{Learning Fair Decisions with Factor Models: Applications to Annuity Pricing}
\author[a]{Fei Huang} 
\author[b]{Junhao Shen} 
\author[b]{Yanrong Yang\thanks{corresponding author: yanrong.yang@anu.edu.au}} 
\author[b]{Ran Zhao\thanks{All coauthors have equal contribution on this project.}}

\affil[a]{UNSW Sydney}
\affil[b]{The Australian National University}

\date{\today}

\maketitle

\begin{abstract}
Fairness-aware statistical learning is essential for mitigating discrimination against protected attributes such as gender, race, and ethnicity in data-driven decision-making. This is particularly critical in high-stakes applications like insurance underwriting and annuity pricing, where biased business decisions can have significant financial and social consequences. Factor models are commonly used in these domains for risk assessment and pricing; however, their predictive outputs may inadvertently introduce or amplify bias. To address this, we propose a Fair Decision Model that incorporates fairness regularization to mitigate outcome disparities. Specifically, the model is designed to ensure that expected decision errors are balanced across demographic groups—a criterion we refer to as Decision Error Parity. We apply this framework to annuity pricing based on mortality modeling. An empirical analysis using Australian mortality data demonstrates that the Fair Decision Model can significantly reduce decision error disparity while also improving predictive accuracy compared to benchmark models, including both traditional and fair factor models.

    
\end{abstract}

\section{Introduction}

The widespread adoption of data-driven decision-making has brought increased scrutiny to fairness in algorithmic models, particularly in high-stakes applications such as credit scoring, criminal justice, hiring, and insurance pricing. In these domains, biased models can entrench or exacerbate existing disparities, making fairness a central concern in machine learning and statistics \citep{barocas2023fairness, mehrabi2021survey, mitchell2021algorithmic, frees-huang-2023-discriminating, charpentier2024insurance}.

Mortality forecasting plays a foundational role in actuarial science, demography, and social policy, and it is central to the pricing of life-contingent financial products such as annuities. A wide range of models has been proposed to capture age-specific and temporal trends. These include the benchmark Lee–Carter model \citep{lee-carter}, Renshaw–Haberman (RH) model \citep{renshaw2006cohort},  the Cairns–Blake–Dowd (CBD) model for older ages \citep{cairns2006two}, age-period-cohort (APC) models \citep{currie2004smoothing}, functional data models \citep{hyndman2007robust}, and extensions like the Plat model \citep{plat2009stochastic} that are popular in pension risk modeling.

Although there is a growing body of literature on fairness and discrimination in insurance –particularly in general insurance contexts \citep{xin2024, lindholm-etal-2022-discrimination, cote2024fair, araiza2022discrimination} –relatively limited attention has been paid to fairness in life-contingent applications, such as mortality forecasting and annuity pricing. This gap is especially significant given the long-term financial implications of such products.

In many jurisdictions, gender---a legally protected attribute---is explicitly and legitimately used in mortality models and the pricing of life-contingent insurance products due to its strong predictive signal for longevity. While most of the fairness literature assumes that protected attributes are not allowed as a rating factor (direct discrimination is not allowed), insurance practice often permits or even requires their inclusion for some lines of business. For example, gender is a commonly used rating factor for insurance in many jurisdictions.  This raises a critical yet underexplored question: \textit{how can we ensure fairness in decision-making when protected attributes are used explicitly as rating factors (direct discrimination is allowed)?} Standard fairness frameworks are not directly applicable in such settings, as they usually focus on how to mitigate indirect discrimination when direct discrimination is prohibited, for example, see \cite{lindholm-etal-2022-discrimination,xin2024}. An innovative contribution of this paper is to address this gap by developing fairness-aware factor models that account for protected attributes when it is allowed to be used as a rating factor in a regression problem. We demonstrate through an annuity pricing example that even when gender is a legitimate rating factor in decision-making and modeling, significant disparities in the accuracy of prediction and decision outcomes can persist, and our proposed methodology is designed to detect and correct these imbalances.

Factor models are a classical and powerful framework for uncovering latent structures in high-dimensional data. They capture systematic variation using a small number of latent components and are widely used in domains such as asset pricing, macroeconomic forecasting, and mortality modeling. In the context of mortality forecasting, factor models—such as the Lee-Carter model \citep{lee-carter} and its variants—have been extensively applied to analyze long-term demographic trends across countries.  The Lee-Carter model, in particular, remains a benchmark in mortality forecasting and is used by the U.S.\ Bureau of the Census for long-run life expectancy projections \citep{hollmann1999methodology}. Beyond the United States, it has been successfully applied in many countries including those in the G7 \citep{tuljapurkar2000universal} as well as Sweden and Australia \citep{lundstrom2004mortality, booth2001age, booth2002applying, booth2003future, booth2004beyond}. 

However, standard factor models are agnostic to fairness considerations. When applied to data containing distinct demographic groups---such as males and females---they may yield imbalanced reconstruction or prediction errors across groups, due
to differences in how well the model captures gender-specific mortality patterns. For example, even when group sizes are equal, the average prediction error for one gender may be substantially higher, which can propagate into downstream decisions such as annuity pricing. These disparities are often invisible in aggregate model performance metrics, yet they raise important concerns in fairness-aware modeling.

Group fairness in machine learning is commonly defined using statistical criteria such as \textit{independence} (also known as demographic parity), \textit{separation} (equalized odds), and \textit{sufficiency} (well-calibration) \citep{barocas2023fairness, Hardt2016EqualityOO, dwork2012fairness, mehrabi2021survey}. These principles have inspired a wide array of algorithmic interventions—spanning pre-processing, in-processing, and post-processing approaches—to mitigate bias in supervised learning \citep{pmlr-v28-zemel13, agarwal2018reductions}. A number of algorithmic techniques have been proposed to introduce fairness constraints into model estimation. For example, methods based on constrained optimization, Pareto optimality, and gradient-based learning have been applied to problems such as subspace learning and dimensionality reduction \citep{samadi2018price, Morgenstern2019FairDR, olfat2019convex, kamani2022efficient, vu2022distributionally, Babu2023FairPC, zalcberg2021fair, kleindessner2023efficient}. These approaches often aim to minimize disparities in model loss across demographic groups, even at the cost of slightly higher overall error.

While promising, many existing fairness-aware methods operate in the predictive machine learning setting without direct links to decision-making applications. In contrast, our work is grounded in the concrete context of insurance pricing, where fairness must be considered not only at the model level but also in the downstream economic decisions informed by model outputs (e.g., annuity pricing). While the fairness notions considered in this paper can be viewed as adaptations of the independence (or demographic parity) criterion to a regression setting, our approach departs from the conventional literature that typically applies this criterion to price outcomes—an approach often infeasible in many insurance lines of business \citep{xin2024, lindholm2024fair}. Instead, we apply the criterion to prediction errors or decision errors, ensuring that accuracy—rather than predicted mortality rates or annuity prices—is balanced across demographic groups such as gender.

Recent work has begun to explore fairness in the context of decision-making policies, highlighting the distinction between fair predictions and fair outcomes. For instance, \citet{Scantamburlo_2024} emphasize that fairness cannot be fully achieved through model constraints alone; the downstream use of predictions must also be considered. Similarly, \citet{shimao2022welfare} show that applying fairness-aware machine learning algorithms on cost modeling in an insurance context cannot achieve fairness in the market price or welfare alone. However, they can significantly harm the insurer’s profit and consumer welfare under certain market conditions, particularly of females.
Our work aligns with this emerging literature by emphasizing fairness in decision-making. We introduce the fairness criterion, \textit{decision error parity}, which targets disparities in downstream decisions rather than prediction accuracy alone. In particular, we focus on annuity pricing as an application, where decisions based on forecasted mortality rates can have significant long-term financial implications.

This paper introduces a fairness-aware framework for learning decisions with factor models. We propose and study three models:
\begin{enumerate}
    \item \textbf{Traditional Factor Model}: the standard baseline model that minimizes overall reconstruction or prediction error without fairness constraints.
    \item \textbf{Fair Factor Model}: a model that incorporates fairness constraints into the factor structure estimation, aiming to reduce disparities in prediction error across demographic groups. This criterion is referred to as Reconstruction Error Parity.
    \item \textbf{Fair Decision Model}: a decision-focused model that directly targets fairness in downstream decision outcomes (e.g., annuity pricing), aiming to reduce disparities in decision errors across demographic groups. This criterion is referred to as Decision Error Parity.
\end{enumerate}

We evaluate our models through an empirical application to Australian mortality data. Our findings show that while traditional factor models introduce substantial gender-based disparities in accuracy, the fair factor model reduces these disparities in mortality forecasting, and the fair decision model further improves accuracy and fairness in annuity pricing outcomes.

\noindent\textbf{Contributions.} This paper makes the following key contributions:

\begin{enumerate}
    \item \textbf{Fairness with protected attributes:} Unlike much of the fairness literature that assumes direct discrimination is not allowed (protected attributes are not allowed as rating factors), we study fairness in settings where such attributes (e.g., gender) are explicitly used and permitted—reflecting many real-world practices in actuarial modeling. We provide a principled framework for defining and achieving fairness in this context, including new criteria tailored to these settings.

    \item \textbf{Fairness in decisions:} We go beyond fairness in machine learning prediction to address fairness in decision-making outcomes, which are functions of the predictive model outputs. By proposing a novel criterion—Decision Error Parity—alongside the Fair Decision Model, we aim to minimize disparities in decision accuracy, particularly in the predictive accuracy of annuity prices, across protected groups.

    \item \textbf{Fairness in factor models for life-contingent insurance applications:} To our knowledge, this is the first study to investigate fairness in the context of factor models applied to mortality forecasting and life-contingent insurance products. We propose two fairness-aware extensions of traditional factor models: the Fair Factor Model, which targets reconstruction (prediction) error parity, and the Fair Decision Model, which targets decision error parity in downstream applications such as annuity pricing. Empirical results based on Australian mortality data show that the Fair Decision Model improves both fairness and predictive accuracy for annuity pricing compared to the two benchmark models.

\end{enumerate}

The remainder of the paper is structured as follows. Section~\ref{sec: motivation} motivates the study by highlighting gender bias in the estimation errors of a standard factor model for mortality modeling. Section~\ref{pre} introduces the traditional factor model, the Fair Factor Model, and their respective estimation methods. Section~\ref{method} presents the Fair Decision Model along with its estimation approach. Section~\ref{real example} provides an empirical analysis using Australian mortality data. The paper concludes with a discussion of broader implications and potential extensions.

\section{Motivation}
\label{sec: motivation}

In mortality modeling and annuity pricing, ensuring prediction accuracy across demographic groups is critical for both fairness and actuarial soundness. While gender is allowed to be used as a rating factor for many jurisdictions and is explicitly included in the modeling process, discrepancies in prediction errors between males and females may still arise due to differences in how well the model captures gender-specific mortality patterns. These discrepancies can lead to biased annuity valuations, with one gender systematically over- or undercharged relative to their true risk. Understanding and addressing these prediction error gaps is therefore essential—not only to improve model performance, but also to ensure equitable outcomes in insurance and pension systems. In this study, we investigate the nature and extent of gender-based prediction error disparities and propose fairness-aware modeling techniques to reduce these differences, thereby enhancing both the accuracy and fairness of mortality and annuity predictions.

\begin{figure}[!h]
    \centering
    \includegraphics[scale=0.5]{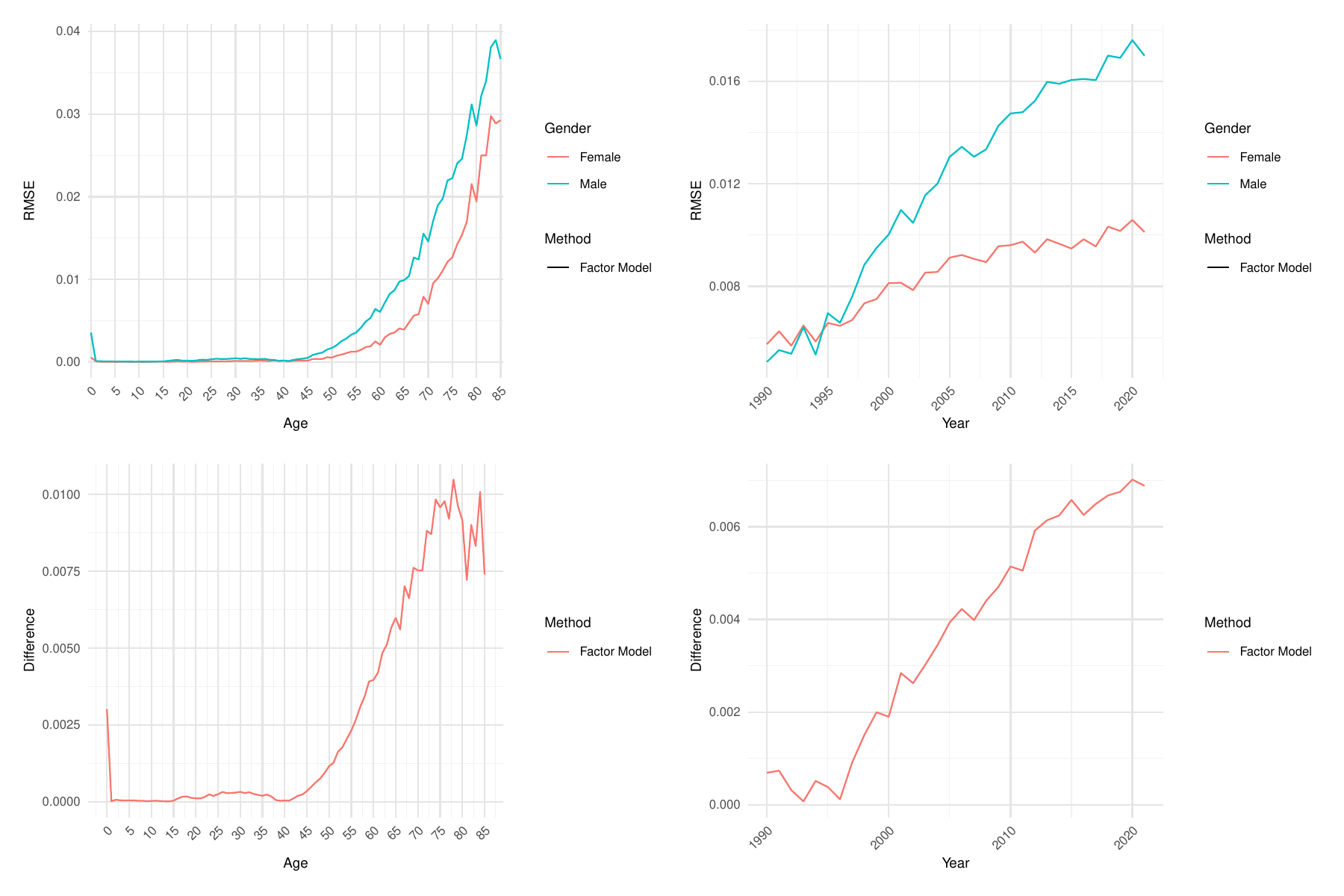}
    \caption{Gender Discrepancies in Mortality Prediction Accuracy Using the Lee-Carter Model.
Top panels: Root Mean Squared Error (RMSE) of estimated mortality rates by gender, age, and year.
Bottom panels: Absolute difference in RMSE between genders by age and year, highlighting the magnitude of prediction accuracy discrepancies. }
    \label{CFM}
\end{figure}

In Figure \ref{CFM}, we illustrate gender disparities in prediction accuracy using the Lee-Carter model for Australian mortality data. The top panels present estimation accuracy, measured by Root Mean Square Error (RMSE). The results indicate that predictive accuracy is higher for females compared to males. The bottom panels highlight notable differences between males and females—particularly for individuals aged 0 and those over 50. These discrepancies become more pronounced with longer-term forecasts, indicating that unfairness tends to intensify over extended prediction horizons.

\begin{figure}[!h]
    \centering
    \includegraphics[scale=0.45]{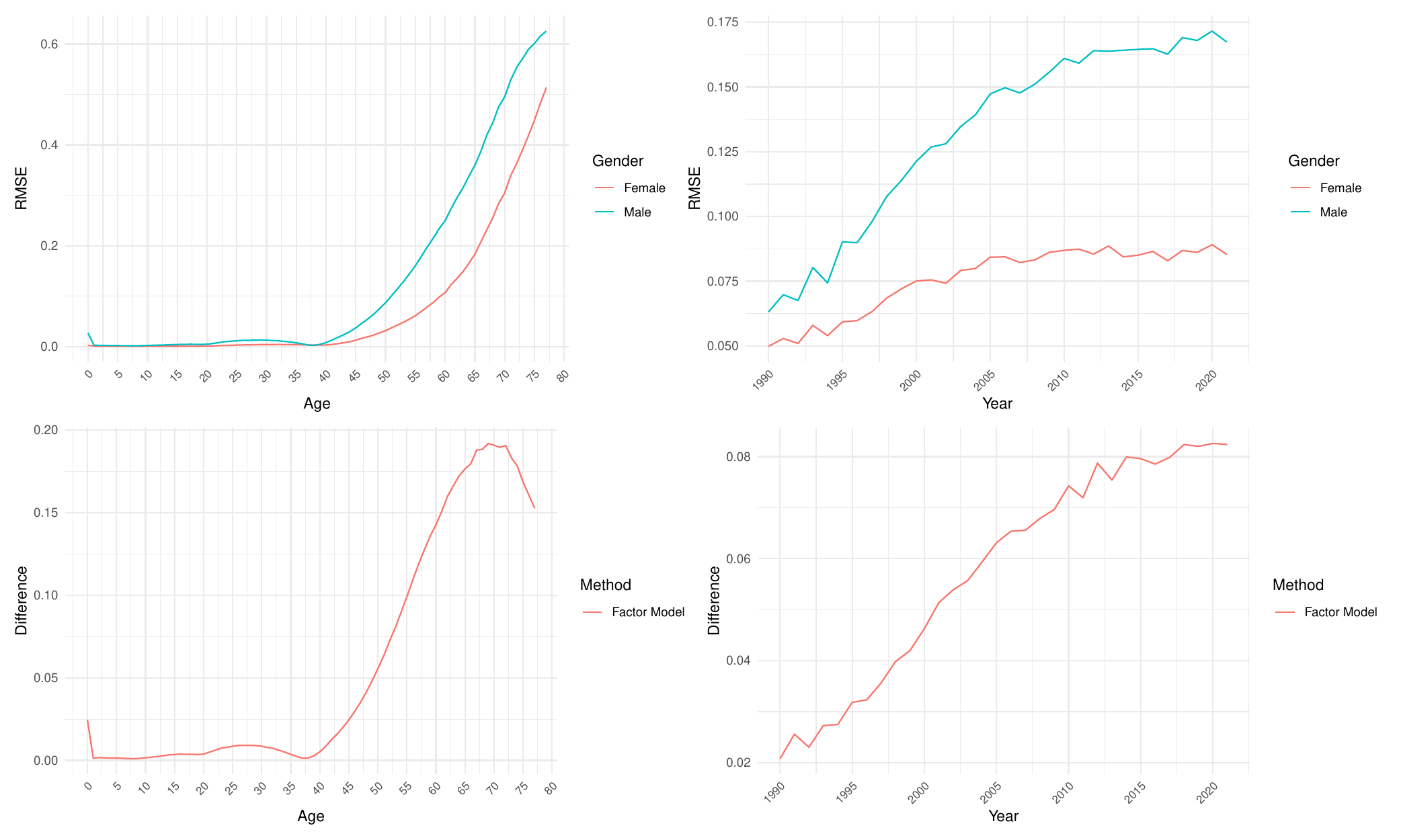}
    \caption{Gender Discrepancies in Annuity Pricing Accuracy Using the Lee-Carter Model.
Top panels:  Root Mean Squared Error (RMSE) of estimated annuity EPVs by gender, age, and year.
Bottom panels: Absolute differences in the RMSE of annuity price estimates between genders across age groups,highlighting the magnitude of prediction accuracy discrepancies.}
    \label{CAFM}
\end{figure}

The expected present value (EPV) of an annuity-due, representing the pure premium (i.e., the actuarially fair value), is a central quantity in annuity pricing and is derived as a function of mortality rates. The top panels of Figure \ref{CAFM} present estimation accuracy of the predicted EPV of an annuity-due using the Lee-Carter model based on the mortality model used in Figure \ref{CFM}. The results show that females exhibit higher predictive accuracy in annuity pricing compared to males. The bottom panels of Figure \ref{CAFM} show substantial gender differences in EPV estimates, especially among older age groups and in longer-range predictions, see Section~\ref{real example} for more modeling details. For an annuity-due with annual payments of \$100,000 over 10 years, the average absolute difference in the RMSE of annuity prices between genders ranges from approximately \$10,000 to around \$20,000 for individuals aged 55 to 80, highlighting a substantial discrepancy in the accuracy of annuity pricing across genders.

To address these issues, this paper aims to mitigate the unfairness arising from estimation error discrepancies in factor models. Our objective is to go beyond predictive fairness and ensure fairness in policy and decision-making—particularly in applications such as annuity pricing, where both accuracy and equity are crucial.

\section{Fair Factor Model}\label{pre}
In this section, we introduce the Fair Factor Model, which extends traditional factor models by incorporating fairness constraints to ensure the resulting factor loading matrix is unbiased. By applying a gradient descent algorithm, the model optimizes both data reconstruction error and fairness, yielding a factor loading matrix that mitigates disparities across sensitive attributes or groups.

We first introduce some notations used throughout this paper. For any vector $\bx=(x_1,\cdots,x_N)^\top$, let $\|\bx\|_2=(\sum_{i=1}^Nx_i^2)^{1/2}$. For a matrix $\Ab$, let $\text{tr}(\Ab)$ denote the trace of $\Ab$ and $\lambda_{j}$ denote the $j$th largest eigenvalue of a nonnegative definite matrix $\Ab$, let $\|\Ab\|_{F}$ denote the Frobenius norm of $\Ab$, $\|\Ab\|_{2}$ be the spectral norm of matrix $\Ab$, and $\|\Ab\|_{\max}$ be the maximum of $|A_{ij}|$. The notation $f(\Ab) = (f(a_{ij}))$ represents a function $f$ applied element-wise to the matrix $\Ab$, where $\Ab$ is a matrix and $a_{ij}$ represents the element in the $i$-th row and $j$-th column of $\Ab$. Let $[T]$ denote the set $\{1,\dots\,T\}$. The notation $\xrightarrow{d}$ represents convergence in distribution. For two random series $X_{n}$ and $Y_{n}$, $X_{n} \lesssim Y_{n}$ means that $X_{n}=O_{p}(Y_{n})$ and $X_{n} \gtrsim Y_{n}$ means that $Y_{n}=O_{p}(X_{n})$. The notation $X_{n} \asymp Y_{n}$ means that $X_{n} \lesssim Y_{n}$ and $X_{n} \gtrsim Y_{n}$. The constants $c$ and $C$ are generic and may represent different values in different expressions.

    \subsection{Factor Model}
Factor models represent high-dimensional time series via latent low-dimensional time series and benefit further inferences. We consider a general factor model allowing unrestricted serial dependence, covering both stationary and nonstationary common factors.    
\begin{equation}\label{fm}
    y_{ti}=c_{ti}+\epsilon_{ti}=\blambda_i^\top\bbf_t+\epsilon_{ti}, 1\leq t \leq T, 1 \leq i \leq N,
\end{equation}
where $c_{ti}=\blambda_i^\top\bbf_t$ is the common component with $\bbf_t \in \RR^r$ being the unobserved common factors and $\blambda_i \in \RR^r$ being the loadings associated with $\bbf_t$, and $\epsilon_{ti}$ is the error component of the factor model.  

The vector form of factor model (\ref{fm}) is 
$$\by_t=\bLambda\bbf_t+\bepsilon_t,$$
where $\by_t^\top=(y_{t1},\cdots,y_{tN}), \bLambda=(\blambda_1,\cdots,\blambda_N)^\top \in \RR^{N \times r}$, and $\bepsilon_t^\top=(\epsilon_{t1},\cdots,\epsilon_{tN})$. However, the latent factors $\bbf_t$ and the factor loading matrix $\bLambda$ in the factor model are not identifiable, because for any invertible matrix $\Hb$, we have $\by_t = \bLambda \bbf_t + \bepsilon_t = (\bLambda \Hb)(\Hb^{-1} \bbf_t)+\bepsilon_t$. To resolve this issue, we impose the following identifiability conditions:
\begin{equation*}
\frac{1}{N}\bLambda^\top\bLambda=\Ib_r, \ \frac{1}{T^{d}}\sum_{t=1}^{T}\bbf_t\bbf_t^\top \ \text{is diagonal}.
\end{equation*}
Here $d\geq 1$ reflects the strength of the time trend in $\{\bbf_t, t=1, \ldots, T\}$. For example, $d=1$ for stationary factors while $d=2$ when common factors follow a unit-root process. 

The matrix form of the factor model is 
$$\mathbf{Y}=\mathbf{F}\bm{\Lambda}^\top+\mathbf{E},$$
where $\mathbf{Y}=(\bm{y}_1,\cdots,\bm{y}_T)^\top \in \RR^{T \times N}, \mathbf{F}=(\bm{f}_1,\cdots, \bm{f}_T)^\top \in \RR^{T \times r}$, and $\mathbf{E}=(\bm{\epsilon}_1,\cdots,\bm{\epsilon}_T)^\top \in \RR^{T \times N}$. It is natural to estimate $\bLambda$ and $\Fb=(\bbf_1,\cdots,\bbf_T^\top)$ by minimizing 
$$\begin{aligned}
    &\min_{\bLambda,\Fb}\dfrac{1}{T}\sum_{t=1}^{T}\|\by_t-\bLambda\bbf_t\|_2^2=\dfrac{1}{T}\|\Yb-\Fb\bLambda^\top\|_F^2\\
    &\text{subject to}\ \dfrac{1}{N}\bLambda^\top\bLambda=\Ib_r \ \text{and}\ \dfrac{1}{T^d}\Fb^\top\Fb\ \text{is diagonal}.
\end{aligned}$$
The basic computation yields $\hat{\bLambda}/\sqrt{N}$ are the eigenvectors corresponding to the top $r$ eigenvalues of the matrix $\Yb^\top \Yb$. Then, $\hat{\Fb}=\Yb\hat{\bLambda}/N$, the common component matrix is estimated by $\hat{\Cb}=\hat{\Fb}\hat{\bLambda}^\top=N^{-1}\Yb\hat{\bLambda}\hat{\bLambda}^\top$. Finally, $\hat{\Eb}=\Yb-\hat{\Fb}\hat{\bLambda}^\top=\Yb-N^{-1}\Yb\hat{\bLambda}\hat{\bLambda}^\top$. Therefore, we can define the reconstruction error for the factor model as follows:
\begin{definition}
    (Reconstruction Error) For any given dadaset $\Yb$ and any loading matrix $\bLambda$, the total reconstruction error of $\Yb$ using $\bLambda$ is defined as:
    \begin{equation}\label{RE}
        \cL(\bLambda)=\dfrac{1}{T}\left\|\Yb-\dfrac{1}{N}\Yb\bLambda\bLambda^\top\right\|_F^2.
    \end{equation}
\end{definition}


\subsection{Fair Factor Model}

The fairness issue arises because the $\bLambda^*$ obtained from minimizing (\ref{RE}) leads to different reconstruction errors for different sensitive groups. While we know that learning a separate subspace for each group results in the optimal reconstruction error, here we consider learning a single optimal loading matrix space in order to balance the reconstruction errors across different groups, taking both statistical and ethical considerations into account. We consider binary attribute $A$ and let $\mathcal{A}=\{1,2\}$. 
We denote the data matrix for each group as $\Yb_k \in \RR^{T_k \times N}$, where $T_k$ is the number of samples in group $k$. 
$$\Yb=\left(\begin{array}{c}
    \Yb_1\\
    \Yb_2
\end{array}\right)=\left(\begin{array}{c}
    \Fb_1\\
    \Fb_2
\end{array}\right)\bLambda^\top+\left(\begin{array}{c}
    \Eb_1 \\
    \Eb_2
\end{array}\right), $$
then, the $\Yb$ is estimated by 
$$\hat{\Yb}=\left(\begin{array}{c}
     \hat{\Yb}_1 \\
     \hat{\Yb}_2
\end{array}\right)=\left(\begin{array}{c}
     \hat{\Fb}_1\hat{\bLambda}^\top  \\
     \hat{\Fb}_2\hat{\bLambda}^\top
\end{array}\right)=\left(\begin{array}{c}
     \Yb_1\hat{\bLambda}\hat{\bLambda}^\top/N  \\
     \Yb_2\hat{\bLambda}\hat{\bLambda}^\top/N
\end{array}\right).$$
The reconstruction error for each group using $\bLambda$ is denoted as $$\cL_k(\bLambda)=\dfrac{1}{T_k}\left\|\Yb_k-\dfrac{1}{N}\Yb_k\bLambda\bLambda^\top\right\|_F^2, 1\leq k \leq 2.$$
Using the definition of (\ref{RE}), we can define a fair factor model as follows.
\begin{definition}
    (Reconstruction Error Parity and Fair Factor Model) A fair factor model with loading matrix $\bLambda$ is called fair, if it satisfies Reconstruction Error Parity, that is the reconstruction errors are equal across different groups, i.e.,
    $$\cL_1(\bLambda)=\cL_2(\bLambda).$$
\end{definition}
This definition is similar to the ``independence'' (or demographic parity) fairness notion in machine learning applied to reconstruction errors. Specifically, ensuring that the disparity error is equal across all groups \citep{kamani2022efficient} can be seen as a relaxation of requiring the reconstruction error to be equal across all groups, satisfying $|\cL_1(\bLambda^*)-\cL_2(\bLambda^*)| \leq \alpha$, where $\alpha=|\cL_1(\bLambda_1^*)-\cL_2(\bLambda_2^*)|$, $\bLambda_1^*=\argmin_{\bLambda} \cL_1(\bLambda)$ and $\bLambda_2^*=\argmin_{\bLambda}\cL_2(\bLambda)$. When both groups can be well represented by a low-dimensional $N\times r$ matrix, $\alpha$ approaches zero.

In this paper, we not only minimize the reconstruction error but also penalize unfairness as part of our objective function. Specifically, for any loading matrix $\bLambda$, we define the unfairness as the squared difference in reconstruction error between the two groups:
$$\mathcal{U}(\bLambda):=\left(\cL_1(\bLambda)-\cL_2(\bLambda)\right)^2.$$

We consider the following fair factor model problem:
\begin{equation}\label{obj}
    \min_{\bm{\Lambda}, \Fb}\dfrac{1}{T}\|\mathbf{Y}-\mathbf{F}\bm{\Lambda}^\top\|_F^2+\lambda\left(\dfrac{1}{T_1}\|\mathbf{Y}_1-\mathbf{F}_1\bm{\Lambda}^\top\|_F^2-\dfrac{1}{T_2}\|\mathbf{Y}_2-\mathbf{F}_2\bm{\Lambda}^\top\|_F^2\right)^2,
\end{equation}
subject to $\dfrac{1}{N}\bLambda^\top\bLambda=\Ib_r$, where $\lambda \geq 0$ is penalty parameter that enforces fairness. Therefore, if $\lambda=0$, it becomes a standard factor model. And when $\lambda$ is larger, this ensures that the factor loading matrix $\bLambda$ satisfies the fairness constraint $\mathcal{U}(\bLambda)=0$. It is important to note that although the penalty term focuses on fairness constraints, both accuracy and fairness can be achieved simultaneously: the fairness regularization not only promotes fairness but also reduces model complexity, leading to a better bias–variance trade-off. This phenomenon can be seen through the empirical analysis in Section \ref{real example}.

For given $\bLambda$, the solutions for the least-square problem is $\hat{\mathbf{F}}=\mathbf{Y}\bLambda/N$, where $\hat{\mathbf{F}}_1=\mathbf{Y}_1\bLambda/N, \hat{\mathbf{F}}_2=\mathbf{Y}_2\bLambda/N$. Substituting them in (\ref{obj}), the objective function now becomes \begin{equation}\label{fair pca}
    \min_{\bLambda^\top\bLambda/N=\Ib_r}\cL(\bLambda)+\lambda(\cL_1(\bLambda)-\cL_2(\bLambda))^2.
\end{equation} Since (\ref{fair pca}) does not have an explicit expression, we can solve it iteratively using the gradient descent method. By computing the gradient of (\ref{fair pca}) with respect to $\bLambda$, we have
\begin{equation}
    \begin{aligned}\label{Gradient}
       \mathcal{G}(\bLambda)=-\dfrac{2}{TN}\mathbf{Y}^\top\mathbf{Y}\bLambda&+4\lambda\left(\dfrac{1}{T_1}\|\mathbf{Y}_1-\dfrac{1}{N}\mathbf{Y}_1\bLambda\bLambda^\top\|_F^2-\dfrac{1}{T_2}\|\mathbf{Y}_2-\dfrac{1}{N}\mathbf{Y}_2\bLambda\bLambda^\top\|_F^2\right)\\
       &\times \left(\dfrac{1}{T_2N}\mathbf{Y}_2^\top\mathbf{Y}_2\bLambda-\dfrac{1}{T_1N}\mathbf{Y}_1^\top\mathbf{Y}_1\bLambda\right).
    \end{aligned}
\end{equation}
The projection can be determined using gradient descent by progressively refining an initial solution through iterative updates:
$$\bLambda_{j+1}=\sqrt{N}\prod_{\mathcal{P}_r}(\bLambda_j-\eta\mathcal{G}(\bLambda_j)),$$
where $\eta$ is the learning rate, which can be selected through backtracking line search or exact line search. And $\prod_{\mathcal{P}_r}$ is the projection operator onto $\mathcal{P}_r=\{\mathbf{U} \in \mathbb{R}^{d \times r}|\mathbf{U}^\top\mathbf{U}=\mathbf{I}_r\}$. The specific iteration process is summarized in Algorithm \ref{alg:1}.

\begin{algorithm}
	\renewcommand{\algorithmicrequire}{\textbf{Input:}}
	\renewcommand{\algorithmicensure}{\textbf{Output:}}
	\caption{The Gradient Descent Algorithm for Fair 
        Factor Model.}
	\label{alg:1}
	\begin{algorithmic}[1]
		\REQUIRE Data matrix $\Yb=\Yb_1 \cup \Yb_2$, low rank $r$, penalty parameter $\lambda$, convergence threshold $\epsilon$, the number of iterations $MaxI$
		\ENSURE Fair factor loading matrix $\hat{\bLambda}$ with $\text{rank}(\bLambda)=r$, factor matrix $\hat{\Fb}_1, \hat{\Fb}_2$
            \STATE $t=0$, draw the initial estimator $\bLambda_0 \in \RR^{d \times r}$ randomly;
            \STATE \textbf{repeat}
            \STATE calculate the gradient $\cG(\bLambda_j)$ in (\ref{Gradient});
		\STATE update $\bLambda_{j+1}$ as 
            $\bLambda_{j+1}=\sqrt{N}\prod_{\cP_r}(\bLambda_j-\eta\cG(\bLambda_j))$ where $\eta$ minimizes equation (\ref{fair pca}) by substituting $\bLambda$ with $\bLambda_{j+1}$;
            \STATE \textbf{until} $\dfrac{\|\Yb\bLambda_{j+1}\bLambda_{j+1}^\top/N-\Yb\bLambda_j\bLambda_j^\top/N\|_F}{\|\Yb\bLambda_j\bLambda_j^\top/N\|_F} \leq \epsilon$ or $j=MaxI$;
		\STATE output the estimator $\hat{\bLambda}$ from the last step and $\hat{\Fb}_1=\Yb_1\hat{\bLambda}/N, \hat{\Fb}_2=\Yb_2\hat{\bLambda}/N$.
	\end{algorithmic}
\end{algorithm}

Now, consider making predictions based on the factor model. According to Algorithm \ref{alg:1}, $\by_t$ is estimated by $\hat{\bLambda}\hat{\bbf_t}$ for $1 \leq t \leq T$. To study the behavior of the prediction, consider another dataset $$\by_{T+h}=\bLambda\bbf_{T+h}+\bepsilon_{T+h}, 1\leq h \leq S,$$ 
which satisfies the same assumptions of model. Hence, $\hat{\bm{y}}_{T+h}=\hat{\bm{\Lambda}}\hat{\bm{f}}_{T+h}, h=1,\cdots, S$, where $\hat{\bm{f}}_{T+h}$ is estimated using $\hat{\bm{f}}_1, \cdots, \hat{\bm{f}}_T$. The error between $\hat{\by}_{T+h}$ and $\by_{T+h}$ is:
$$\begin{aligned}
    \|\hat{\bm{y}}_{T+h}-\bm{y}_{T+h}\|_2^2&=\|\hat{\bm{\Lambda}}\hat{\bm{f}}_{T+h}-\bm{\Lambda}\bm{f}_{T+h}-\bm{\epsilon}_{T+h}\|_2^2\\
    &=\|\hat{\bm{\Lambda}}\hat{\bm{f}}_{T+h}-\hat{\bm{\Lambda}}\bm{f}_{T+h}+\hat{\bm{\Lambda}}\bm{f}_{T+h}-\bm{\Lambda}\bm{f}_{T+h}-\bm{\epsilon}_{T+h}\|_2^2\\
    &=\|\hat{\bm{\Lambda}}(\hat{\bm{f}}_{T+h}-\bm{f}_{T+h})+(\hat{\bm{\Lambda}}-\bm{\Lambda})\bm{f}_{T+h}-\bm{\epsilon}_{T+h}\|_2^2\\
    &\leq 2\|\hat{\bm{\Lambda}}\|_F^2\|\hat{\bm{f}}_{T+h}-\bm{f}_{T+h}\|_2^2+\|\hat{\bm{\Lambda}}-\bm{\Lambda}\|_F^2\|\bm{f}_{T+h}\|_2^2+\|\bm{\epsilon}_{T+h}\|_2^2.
\end{aligned}$$
If the reconstruction error is independent of sensitive attributes such as gender, and we can consistently estimate the latent factors and factor loading matrix, then the leading term will be $\bm{\epsilon}_{T+h}$. Consequently, the prediction error will also be independent of gender, defined as \textit{Predictive Error Parity}.

Given $\hat{\mathbf{F}}_1^\top=(\hat{\bm{f}}_1^{(1)}, \cdots, \hat{\bm{f}}_{T_1}^{(1)})$ and $\hat{\mathbf{F}}_2^\top=(\hat{\bm{f}}_1^{(2)},\cdots,\hat{\bm{f}}_{T_2}^{(2)})$, we can estimate $\hat{\bm{f}}_{T+h_1}^{(1)}$ and $\hat{\bbf}_{T+h_2}^{(2)}$ for $h_1=1, \cdots, S_1$ and $h_2=1,\cdots,S_2$, respectively. Thus, we have
$\hat{\bm{y}}_{T+h_1}^{(1)}=\hat{\bm{\Lambda}}\hat{\bm{f}}_{T+h_1}^{(1)}, \hat{\bm{y}}_{T+h_2}^{(2)}=\hat{\bm{\Lambda}}\hat{\bm{f}}_{T+h_2}^{(2)}$. Given the sample, we have the following result of prediction error.
$$\dfrac{1}{S_1}\sum_{h_1=1}^{S_1}\|\hat{\by}_{T+h_1}^{(1)}-\bm{y}_{T+h_1}^{(1)}\|_2^2 \approx\dfrac{1}{S_2}\sum_{h_2=1}^{S_2}\|\hat{\bm{y}}_{T+h_2}^{(2)}-\bm{y}_{T+h_2}^{(2)}\|_2^2.$$

\subsection{Multiclass Fair Factor Model}
Although we focus on binary sensitive attributes in this paper, it is straightforward to extend the methodology to multiclass attributes where $\mathcal{A}=\{1, \cdots,K\}$. In this setting, we aim to find a $\bLambda$ such that $\cL_k(\bLambda)=\cL_{k^\prime}(\bLambda)$ for all $k < k^\prime$. Let 
$\Yb_k \in \RR^{T_k \times N}$ denote the data matrix corresponding to group $k$, where $T_k$ represents the sample size of group $k, k \in \{1, \cdots, K\}$. Let $\Fb_k \in \RR^{T_k \times r}$ denote the factor matrix for the $k$-th group. Let $T$ denote the total sample size across all groups, where $T = T_1 + \cdots + T_K$. We consider the following minimization problem:
$$\min_{\bLambda,\Fb_1, \cdots, \Fb_K}\dfrac{1}{T}\sum_{k=1}^K\|\Yb_k-\Fb_k\bLambda^\top\|_F^2+\lambda\sum_{k=1}^K\sum_{k < k^\prime}^K\left(\dfrac{1}{T_k}\|\Yb_k-\Fb_k\bLambda^\top\|_F^2-\dfrac{1}{T_{k^\prime}}\|\Yb_{k^\prime}-\Fb_{k^\prime}\bLambda^\top\|_F^2\right)^2.$$
By substituting $\Fb_1=\Yb_1\bLambda/N, \cdots, \Fb_K=\Yb_K\bLambda/N$ into the objective, we obtain
$$\min_{\bLambda^\top\bLambda=\Ib_r}\cL(\bLambda)+\lambda\sum_{k=1}^K\sum_{k < k^\prime}^K\left(\cL_k(\bLambda)-\cL_{k^\prime}(\bLambda)\right)^2.$$
By computing the gradient with respect to $\bLambda$, the solution can be obtained iteratively using a gradient descent method similar to Algorithm \ref{alg:1}. Therefore, we omit the details here.

\subsection{Mortality Modeling}

We can naturally apply the Factor Model and the Fair Factor Model to mortality modeling. Assume that gender (male and female) is the protected variable. A standard Factor Model can be used to fit the matrix of mortality rates, as originally proposed in \cite{lee-carter}. Specifically, the mortality rate for age $i$ in year $t$ is modeled as $m_{ti} = \exp(y_{ti} + a_i)$, where $y_{ti}$ follows the factor model specified in (\ref{fm}), and $a_i$ represents the age-specific intercept term. Therefore, $$ \ln(m_{ti})=a_i+\bm{\lambda}_i^\top\bm{f}_t+\epsilon_{ti}, \quad \text{or} \quad \ln(\bm{m}_t)=\bm{a}+\bm{\Lambda}\bm{f}_t+\bm{\epsilon}_t,$$
where $\bm{a}$ is the intercept term. We model $\ln(\bm{m}_t)-\bm{a}$ and denote it as $\bm{y}_t$. We have $\bbm_t=\exp(\by_t+\ba)$. 

By applying the Fair Factor Model to estimate the loading matrix and factors, we obtain $$\hat{\bbm_t}=\exp(\hat{\bLambda}\hat{\bbf}_t+\hat{\ba}).$$ In practice, $\hat{\ba}$ is estimated as the simple average of $\ln(\bm{m}_t)$ over time. While traditional mortality modeling literature typically ignores potential estimation bias between genders, the Fair Factor Model allows us to explicitly account for such disparities.

\section{Fair Decision Model}\label{method}
In this section, we propose a \emph{fair decision model} that directly targets fairness in the decision-making stage, along with a gradient descent algorithm for its estimation. By incorporating fairness constraints into the objective function, our approach can enable both fair and accurate decision-making in practical applications.

In many data-driven decision support systems, factor models are commonly used as a preprocessing step. The typical pipeline involves two stages: (i) extracting low-dimensional latent features from high-dimensional data using a factor model, and (ii) using the extracted features for downstream tasks such as prediction, modeling, or policy decisions (such as annuity pricing). While traditional approaches focus on minimizing reconstruction error in the first stage, fairness considerations are often overlooked in the second stage where decisions are made.

To address this gap, we extend our framework beyond the factor estimation stage and directly analyze disparities in decision outcomes. We introduce a formal definition of the \emph{fair decision model}, which aims to ensure that decision errors are balanced across demographic groups.

\begin{definition}
    (Decision Error Parity and Fair Decision Model) The decision model with loading matrix $\bLambda$ is fair if Decision Error Parity is satisfied, that is the decision errors are equal across different groups,
    $$\dfrac{1}{T_1}\left\|g\left(\dfrac{1}{N}\Yb_1\bLambda\bLambda^\top\right)-g\left(\Yb_1\right)\right\|_F^2=\dfrac{1}{T_2}\left\|g\left(\dfrac{1}{N}\Yb_2\bLambda\bLambda^\top\right)-g\left(\Yb_2\right)\right\|_F^2,$$
    where $g(\cdot)$ is a certain functional transformation, $\Yb_1=\Fb_1\bLambda^\top+\Eb_1$ and $\Yb_2=\Fb_2\bLambda^\top+\Eb_2$. 
\end{definition}

We then propose the following fair decision model, 
\begin{equation}\label{fair decision model}
    \min_{\bLambda,\Fb}\dfrac{1}{T}\left\|g(\Fb\bLambda^\top)-g(\Yb)\right\|_F^2+\lambda\left(\dfrac{1}{T_1}\left\|g(\Fb_1\bLambda^\top)-g(\Yb_1)\right\|_F^2-\dfrac{1}{T_2}\left\|g(\Fb_2\bLambda^\top)-g(\Yb_2)\right\|_F^2\right)^2.
\end{equation}
(\ref{fair decision model}) can encompass a general fair factor model and can also be applied to other fields. When $g(x)=x$, fair decision model (\ref{fair decision model}) reduces to the fair factor model (\ref{obj}). Another example can be found in subsection \ref{FPDM}.

Substituting $\hat{\mathbf{F}}=\mathbf{Y}\bLambda/N$, where $\hat{\mathbf{F}}_1=\mathbf{Y}_1\bLambda/N, \hat{\mathbf{F}}_2=\mathbf{Y}_2\bLambda/N$ in (\ref{fair decision model}), and we use 
$$\begin{aligned}
    x&=\left\|g\left(\Yb\bLambda\bLambda^\top/N\right)-g(\Yb)\right\|_F^2=\sum_{t=1}^T\left\|g\left(\bLambda\bLambda^\top\by_t/N\right)-g(\by_t)\right\|_2^2\\
    &=\sum_{t=1}^T\left[g\left(\bLambda\bLambda^\top\by_t/N\right)-g(\by_t)\right]^\top\left[g\left(\bLambda\bLambda^\top\by_t/N\right)-g(\by_t)\right]
\end{aligned}$$ as an example, its derivative is given by:
$$\begin{aligned}
    \dfrac{\partial x}{\partial \bLambda}&=\dfrac{2}{N}\sum_{t=1}^T\text{diag}(g^\prime(\bLambda\bLambda^\top\by_t/N))\left[g(\bLambda\bLambda^\top\by_t/N)-f(\by_t)\right]\by_t^\top\bLambda\\
    &+\dfrac{2}{N}\sum_{t=1}^T\by_t\left[g(\bLambda\bLambda^\top\by_t/N)-g(\by_t)\right]^\top\text{diag}(g^\prime(\bLambda\bLambda^\top\by_t/N))\bLambda.  
\end{aligned}$$

The derivative of (\ref{fair decision model}) with respect to $\bLambda$ can be written as
\begin{equation}\label{G_FDM}
    \begin{aligned}
    \cG(\bLambda)=&\dfrac{2}{TN}\sum_{t=1}^T\text{diag}(g^\prime(\bLambda\bLambda^\top\by_t/N))\left[g(\bLambda\bLambda^\top
    \by_t/N)-g(\by_t)\right]\by_t^\top\bLambda\\
    &+\dfrac{2}{TN}\sum_{t=1}^T\by_t\left[g(\bLambda\bLambda^\top\by_t/N)-g(\by_t)\right]^\top\text{diag}(g^\prime(\bLambda\bLambda^\top\by_t/N))\bLambda\\
    &+4\lambda\left(\dfrac{1}{T_1}\left\|g(\Yb_1\bLambda\bLambda^\top/N)-g(\Yb_1)\right\|_F^2-\dfrac{1}{T_2}\left\|g(\Yb_2\bLambda\bLambda^\top/N)-g(\Yb_2)\right\|_F^2\right)\\
    &\times \left(\dfrac{1}{T_1N}\sum_{t=1}^{T_1}\text{diag}(g^\prime(\bLambda\bLambda^\top\by_t^{(1)}/N))\left[g(\bLambda\bLambda^\top\by_t^{(1)}/N)-g(\by_t^{(1)})\right]\by_t^{(1)\top}\bLambda\right.\\
    &\left. +\dfrac{1}{T_1N}\sum_{t=1}^{T_1}\by_t^{(1)}\left[g(\bLambda\bLambda^\top\by_t^{(1)}/N)-g(\by_t^{(1)})\right]^\top
    \text{diag}(g^\prime(\bLambda\bLambda^\top\by_t^{(1)}/N))\bLambda\right.\\
    &\left. -\dfrac{1}{T_2N}\sum_{t=1}^{T_2}\text{diag}(g^\prime(\bLambda\bLambda^\top\by_t^{(2)}/N))\left[g(\bLambda\bLambda^\top\by_t^{(2)}/N)-g(\by_t^{(2)})\right]\by_t^{(2)\top}\bLambda\right.\\
    &\left. -\dfrac{1}{T_2N}\sum_{t=1}^{T_2}\by_t^{(2)}\left[g(\bLambda\bLambda^\top\by_t^{(2)}/N)-g(\by_t^{(2)})\right]^\top
    \text{diag}(g^\prime(\bLambda\bLambda^\top\by_t^{(2)}/N))\bLambda\right).
\end{aligned}
\end{equation}

The projection can be determined using gradient descent by progressively refining an initial solution through iterative updates:
$$\bLambda_{j+1}=\sqrt{N}\prod_{\mathcal{P}_r}(\bLambda_j-\eta\mathcal{G}(\bLambda_j)).$$ The specific iteration process is summarized in Algorithm \ref{alg:2}.

\begin{algorithm}
	\renewcommand{\algorithmicrequire}{\textbf{Input:}}
	\renewcommand{\algorithmicensure}{\textbf{Output:}}
	\caption{The Gradient Descent Algorithm for Fair Decision Model.}
	\label{alg:2}
	\begin{algorithmic}[1]
		\REQUIRE Data matrix $\Yb=\Yb_1 \cup \Yb_2$, function $g$, low rank $r$, penalty parameter $\lambda$, convergence threshold $\epsilon$, the number of iterations $MaxI$
		\ENSURE Fair factor loading matrix $\bLambda$ with $\text{rank}(\bLambda)=r$, factor matrix $\hat{\Fb}_1, \hat{\Fb}_2$, estimators $g(\hat{\Yb}_1), g(\hat{\Yb}_2)$
            \STATE $t=0$, draw the initial estimator $\bLambda_0 \in \RR^{N \times r}$ randomly;
            \STATE \textbf{repeat}
            \STATE calculate the gradient $\cG(\bLambda_j)$ in (\ref{G_FDM});
		\STATE update $\bLambda_{j+1}$ as $\bLambda_{j+1}=\sqrt{N}\prod_{\cP_r}(\bLambda_j-\eta\cG(\bLambda_j))$, where $\eta$ minimizes equation (\ref{fair decision model}) by substituting $\bLambda$ with $\bLambda_{j+1}$;
            \STATE \textbf{until} $\dfrac{\|\left(g(\Yb\bLambda_{j+1}\bLambda_{j+1}^\top/N)-g(\Yb\bLambda_j\bLambda_j^\top/N)\right)\|_F}{\|g(\Yb\bLambda_j\bLambda_j^\top/N)\|_F} \leq \epsilon$ or $j=MaxI$;
		\STATE output the estimator $\hat{\bLambda}$ from the last step, $\hat{\Fb}_1=\Yb_1\hat{\bLambda}/N, \hat{\Fb}_2=\Yb_2\hat{\bLambda}/N$, and $g(\Yb_1\hat{\bLambda}\hat{\bLambda}^\top/N), g(\Yb_2\hat{\bLambda}\hat{\bLambda}^\top/N)$.
	\end{algorithmic}
\end{algorithm}

\subsection{Multiclass Fair Decision Model}
We briefly introduce the multiclass Fair Decision Model when the protected variable has more than two categories, which closely resembles the multiclass Fair Factor Model, except that a functional transformation is applied to the factor model. In the multiclass case, we assume the attribute set $\mathcal{A}=\{1,\cdots,K\}$, and aim to obtain $\bLambda$ such that 
$$\dfrac{1}{T_k}\left\|g\left(\dfrac{1}{N}\Yb_k\bLambda\bLambda^\top\right)-g\left(\Yb_k\right)\right\|_F^2=\dfrac{1}{T_{k^\prime}}\left\|g\left(\dfrac{1}{N}\Yb_{k^\prime}\bLambda\bLambda^\top\right)-g\left(\Yb_{k^\prime}\right)\right\|_F^2$$
for all $k < k^\prime$. Then, the multiclass Fair Decision Model can be represented as follows:
\begin{footnotesize}
    $$\min_{\bLambda,\Fb_1, \cdots, \Fb_K}\dfrac{1}{T}\sum_{k=1}^K\left\|g\left(\Fb_k\bLambda^\top\right)-g\left(\Yb_k\right)\right\|_F^2+\lambda\sum_{k=1}^K\sum_{k < k^\prime}^K\left(\dfrac{1}{T_k}\left\|g\left(\Fb_k\bLambda^\top\right)-g\left(\Yb_k\right)\right\|_F^2-\dfrac{1}{T_{k^\prime}}\left\|g\left(\Fb_{k^\prime}\bLambda^\top\right)-g\left(\Yb_{k^\prime}\right)\right\|_F^2\right)^2.$$
\end{footnotesize}
Substituting $\Fb_1=\Yb_1\bLambda/N, \cdots, \Fb_K=\Yb_K\bLambda/N$ into the above objective function yields:
$$\begin{aligned}
    \min_{\bLambda}&\dfrac{1}{T}\sum_{k=1}^K\left\|g\left(\dfrac{1}{N}\Yb_k\bLambda\bLambda^\top\right)-g\left(\Yb_k\right)\right\|_F^2\\
    &+\sum_{k=1}^K\sum_{k < k^\prime}^K\lambda\left(\dfrac{1}{T_k}\left\|g\left(\dfrac{1}{N}\Yb_k\bLambda\bLambda^\top\right)-g\left(\Yb_k\right)\right\|_F^2-\dfrac{1}{T_{k^\prime}}\left\|g\left(\dfrac{1}{N}\Yb_{k^\prime}\bLambda\bLambda^\top\right)-g\left(\Yb_{k^\prime}\right)\right\|_F^2\right)^2.
\end{aligned}$$
The loading matrix $\bLambda$ is estimated iteratively using a gradient descent method. Once $\bLambda$ is obtained, the estimators of $\Fb_1, \dots, \Fb_K$ follow directly. The algorithm for the multiclass Fair Decision Model closely resembles Algorithm \ref{alg:2}; therefore, we omit the details here.

\subsection{Fair Decision Model for Annuity Pricing}\label{FPDM}
The expected present value (EPV) of an annuity-due represents the pure premium—i.e., the actuarially fair value of the annuity—making it a central quantity in annuity pricing. When we consider the EPV of an annuity-due, it is given by $g(y_{ti})=\sum_{s=0}^{n-1}v^s\prod_{k=0}^{s-1}(1-\exp(y_{t,i+k}+a_{i+k}))$, where $v$ is the discount factor per year, $n$ denotes the term of the annuity, $\exp(y_{ti}+a_i)$ is the death rate at age $i$ in year $t$, denoted as $m_{ti}$. Each $g(y_{ti})$ is a function of $\by_t$, and thus also a function of $\bbm_t$. For convenience, we further denote $g(y_{ti})$ as $p_{ti}(\bbm_t)=\sum_{s=0}^{n-1}v^s\prod_{k=0}^{s-1}(1-m_{t,i+k})$. According to (\ref{fair decision model}), the fair pricing decision model can be expressed as 
    $$\begin{aligned}
        \min_{\bLambda, \Fb}&\dfrac{1}{T}\sum_{t=1}^T\sum_{i=1}^{N-n+2}\left(p_{ti}\left(\exp\left(\bLambda\bbf_t+\ba\right)\right)-p_{ti}\left(\exp\left(\by_t+\ba\right)\right)\right)^2\\
        &+\lambda\left(\dfrac{1}{T_1}\sum_{t=1}^{T_1}\sum_{i=1}^{(N-n+2)}\left(p_{ti}\left(\exp\left(\bLambda\bbf_t^{(1)}+\ba^{(1)}\right)\right)-p_{ti}\left(\exp\left(\by_t^{(1)}+\ba^{(1)}\right)\right)\right)^2\right.\\
        &\left.-\dfrac{1}{T_2}\sum_{t=1}^{T_2}\sum_{i=1}^{N-n+2}\left(p_{ti}\left(\exp\left(\bLambda\bbf_t^{(2)}+\ba^{(2)}\right)\right)-p_{ti}\left(\exp\left(\by_t^{(2)}+\ba^{(2)}\right)\right)\right)^2\right)^2.
    \end{aligned}$$
    Here, $\ba$ denotes the intercept term. Specifically, $\ba^{(1)}$ represents the intercept for the male subgroup ($A = 1$), and $\ba^{(2)}$ represents the intercept for the female subgroup ($A =2$). In this subsection, we focus on the binary case, considering only female and male groups.


    
    Denote $\bbm_t=\exp(\by_t+\ba), \tilde{\bbm}_t=\exp(\bLambda\bbf_t+\ba), \bbm_t^{(1)}=\exp(\by_t^{(1)}+\ba^{(1)}), \tilde{\bbm}_t^{(1)}=\exp(\bLambda\bbf_t^{(1)}+\ba^{(1)}), \bbm_t^{(2)}=\exp(\by_t^{(2)}+\ba^{(2)}), \tilde{\bbm}_t^{(2)}=\exp(\bLambda\bbf_t^{(2)}+\ba^{(2)})$. Using a Taylor expansion, we have:
$$p_{it}(\tilde{\bbm}_t)=p_{it}(\bbm_t)+p_{it}^\prime(\bbm_t)^\top(\tilde{\bbm}_t-\bbm_t)+o_p(1),$$
where 
$$\begin{aligned}    p_{it}^\prime(\bbm_t)&=\left(\underbrace{0,\cdots,0}_{i-1},-\sum_{s=1}^{n-1}v^s\prod_{k=1}^{s-1}(1-m_{i+k,t}),-\sum_{s=2}^{n-1}v^s\prod_{k=0.k\neq 1}^{s-1}(1-m_{i+k,t}),\right.\\
    &\left.-\sum_{s=3}^{n-1}v^s\prod_{k=0,k\neq 2}^{s-1}(1-m_{i+k,t}),\cdots,-v^{n-1}\prod_{k=0}^{n-3}(1-m_{i+k,t}),\underbrace{0,\cdots,0}_{N-i-n+2}\right)^\top \in \RR^N.
\end{aligned}$$

Let $\Wb_t^\top=(p_{1t}^\prime(\bbm_t),p_{2t}^\prime(\bbm_t),\cdots,p_{(N-n+2)t}^\prime(\bbm_t)) \in \RR^{N \times (N-n+2)}$, then $\sum_{i=1}^{N-n+2}(p_{it}^\prime(\bbm_t)^\top(\tilde{\bbm}_t-\bbm_t))^2=\|\Wb_t(\tilde{\bbm}_t-\bbm_t)\|_2^2$. Therefore, optimizing the fair pricing decision model reduces to solving the optimization problem as follows
\begin{equation}\label{com_obj}
    \begin{aligned}
        \min_{\bLambda, \Fb}&\dfrac{1}{T}\sum_{t=1}^T\|\Wb_t\left[\exp(\bLambda\bbf_t+\ba)-\exp(\by_t+\ba)\right]\|_2^2\\
        &+\lambda\left(\dfrac{1}{T_1}\sum_{t=1}^{T_1}\left\|\Wb_t^{(1)}\left[\exp(\bLambda\bbf_t^{(1)}+\ba^{(1)})-\exp(\by_t^{(1)}+\ba^{(1)})\right]\right\|_2^2\right.\\
    &\left.-\dfrac{1}{T_2}\sum_{t=1}^{T_2}\left\|\Wb_t^{(2)}\left[\exp(\bLambda\bbf_t^{(2)}+\ba^{(2)})-\exp(\by_t^{(2)}+\ba^{(2)})\right]\right\|_2^2\right)^2,
\end{aligned}
\end{equation}
where $\Wb_{t}^{(1)}$ and $\Wb_{t}^{(2)}$ are the weights calculated based on subgroup 1 and subgroup 2, respectively. 


Substituting $\hat{\mathbf{F}}=\mathbf{Y}\bLambda/N$, where $\hat{\mathbf{F}}_1=\mathbf{Y}_1\bLambda/N, \hat{\mathbf{F}}_2=\mathbf{Y}_2\bLambda/N$ in (\ref{com_obj}), then the  derivative of (\ref{com_obj}) with respect to $\bLambda$ can be written as
\begin{footnotesize}
\begin{equation}\label{G_insurance}
    \begin{aligned}
    &\cG(\bLambda)=\dfrac{2}{TN}\sum_{t=1}^T\text{diag}(\exp(\bLambda\bLambda^\top\by_t/N+\ba))\Wb_t^\top\Wb_t\left[\exp(\bLambda\bLambda^\top
    \by_t/N+\ba)-\exp(\by_t+\ba)\right]\by_t^\top\bLambda\\
    &+\dfrac{2}{TN}\sum_{t=1}^T\by_t\left[\exp(\bLambda\bLambda^\top\by_t/N+\ba)-\exp(\by_t+\ba)\right]^\top\Wb_t^\top\Wb_t\text{diag}(\exp(\bLambda\bLambda^\top\by_t/N+\ba))\bLambda\\
    &+4\lambda\left(\dfrac{1}{T_1}\sum_{t=1}^{T_1}\left\|\Wb_t^{(1)}\left[\exp(\bLambda\bLambda^\top\by_t^{(1)}/N+\ba^{(1)})-\exp(\by_t^{(1)}+\ba^{(1)})\right]\right\|_2^2\right.\\
    &\left.-\dfrac{1}{T_2}\sum_{t=1}^{T_2}\left\|\Wb_t^{(2)}\left[\exp(\bLambda\bLambda^\top\by_t^{(2)}/N+\ba^{(2)})-\exp(\by_t^{(2)}+\ba^{(2)})\right]\right\|_2^2\right)\\
    &\times \left(\dfrac{1}{T_1N}\sum_{t=1}^{T_1}\text{diag}(\exp(\bLambda\bLambda^\top\by_t^{(1)}/N+\ba^{(1)}))\Wb_t^{(1)\top}\Wb_t^{(1)}\left[\exp(\bLambda\bLambda^\top\by_t^{(1)}/N+\ba^{(1)})-\exp(\by_t^{(1)}+\ba^{(1)})\right]\by_t^{(1)\top}\bLambda\right.\\
    &\left. +\dfrac{1}{T_1N}\sum_{t=1}^{T_1}\by_t^{(1)}\left[\exp(\bLambda\bLambda^\top\by_t^{(1)}/N+\ba)^{(1)}-\exp(\by_t^{(1)}+\ba^{(1)})\right]^\top
    \Wb_t^{(1)\top}\Wb_t^{(1)}\text{diag}(\exp(\bLambda\bLambda^\top\by_t^{(1)}/N+\ba^{(1)}))\bLambda\right.\\
    &\left. -\dfrac{1}{T_2N}\sum_{t=1}^{T_2}\text{diag}(\exp(\bLambda\bLambda^\top\by_t^{(2)}/N+\ba^{(2)}))\Wb_t^{(2)\top}\Wb_t^{(2)}\left[\exp(\bLambda\bLambda^\top\by_t^{(2)}/N+\ba^{(2)})-\exp(\by_t^{(2)}+\ba^{(2)})\right]\by_t^{(2)\top}\bLambda\right.\\
    &\left. -\dfrac{1}{T_2N}\sum_{t=1}^{T_2}\by_t^{(2)}\left[\exp(\bLambda\bLambda^\top\by_t^{(2)}/N+\ba)^{(2)}-\exp(\by_t^{(2)}+\ba^{(2)})\right]^\top
    \Wb_t^{(2)\top}\Wb_t^{(2)}\text{diag}(\exp(\bLambda\bLambda^\top\by_t^{(2)}/N+\ba^{(2)}))\bLambda\right).
\end{aligned}
\end{equation}
\end{footnotesize}
The specific iteration process is similar to Algorithm \ref{alg:2}.


\subsection{Choice of $\lambda$}
The penalty parameter, $\lambda$, controls the trade-off between fairness and accuracy. 
To select an appropriate penalty parameter $\lambda$, we employ $k$-fold cross-validation. The original samples $\Yb^\top=(\Yb_1^\top,\Yb_2^\top)$ are randomly divided into $k$ approximately equal subsets, denoted as $\Yb_{1j}\in \RR^{T_{1j} \times N}$ and $\Yb_{2j}\in \RR^{T_{2j} \times N}$. For each of the $k$ folds, a single subset $\Yb_{j}^\top=(\Yb_{1j}^\top,\Yb_{2j}^\top)$ is used as the validation set, while the remaining subsets serve as the training set. The cross-validation process is repeated $k$ times, with each subset being used exactly once as the validation set. The prediction error for $k$-fold cross-validation is calculated as follows:
$$\begin{small}
    \begin{aligned}
    CV_\lambda&=\dfrac{1}{k}\sum_{j=1}^k\left[\dfrac{1}{T_{1j}+T_{2j}}\left\|g\left(\dfrac{1}{N}\Yb_{j}\hat{\bLambda}_{-j}\hat{\bLambda}_{-j}^\top\right)-g(\Yb_{j})\right\|_F^2\right]\\
    &\text{s.t.} \dfrac{1}{k}\sum_{j=1}^k\left|\dfrac{1}{T_{1j}}\left\|g\left(\dfrac{1}{N}\Yb_{1j}\hat{\bLambda}_{-j}\hat{\bLambda}_{-j}^\top\right)-g(\Yb_{1j})\right\|_F^2-\dfrac{1}{T_{2j}}\left\|g\left(\dfrac{1}{N}\Yb_{2j}\hat{\bLambda}_{-j}\hat{\bLambda}_{-j}\right)-g(\Yb_{2j})\right\|_F^2\right| \leq \lambda_c,
\end{aligned}
\end{small}$$
where $\lambda_c$ is a predetermined constant threshold; $\hat{\bLambda}_{-j}$ is the fair subspace;
$T_{1j}$ and $T_{2j}$ are the sample sizes of the two subsets, respectively. We would like to choose $\lambda$ which minimizes the prediction error $CV_\lambda$ under the fairness requirement. The choice of $\lambda$ in the multiclass case follows a similar approach to that of the binary case.

\section{Empirical Analysis}\label{real example}
\subsection{Fair Mortality Forecasting}\label{Mortality Application}
In this section, we analyze Australian mortality data obtained from the Human Mortality Database (HMD), a comprehensive source of life expectancy and mortality statistics that reports age-specific death rates separately for males and females \citep{hmd2024}. The dataset includes period death rates from 1921 to 2021, with both age and time recorded at annual intervals (1×1 format). Our analysis focuses on ages 0 to 85. To stabilize variance and address potential non-zero means, we apply a logarithmic transformation followed by standardization.




In the data preprocessing stage, mortality data from 1989 and earlier were used as training data for analysis and prediction, while data from 1990 onward were reserved for evaluating predictive performance. This evaluation involved comparing observed mortality rates with those predicted by different techniques: the Factor Model, the Fair Factor Model, and the Fair Decision Model. To capture temporal patterns, an ARIMA model was estimated (based on the corrected Akaike Information Criterion, AICc) separately for each gender and applied to forecast male and female mortality rates.

To evaluate the model performance, we use the root mean squared error (RMSE) to quantify  \textit{accuracy} and the absolute difference of RMSE between genders to quantify \textit{fairness}. 

The accuracy measures for each gender and the overall data are defined below. 
$$\text{RMSE (male)}=\text{RMSE}_1=\sqrt{\dfrac{1}{T_1N}\sum_{t=1}^{T_1}\sum_{i=1}^N(m_{ti}^{(1)}-\hat{m}_{ti}^{(1)})^2}$$ 
$$\text{RMSE
 (female)}=\text{RMSE}_2=\sqrt{\dfrac{1}{T_2N}\sum_{t=1}^{T_2}\sum_{i=1}^N(m_{ti}^{(2)}-\hat{m}_{ti}^{(2)})^2},$$
$$\text{RMSE (total)}=\sqrt{\dfrac{1}{TN}\sum_{t=1}^T\sum_{i=1}^N(m_{ti}-\hat{m}_{ti})^2}=\sqrt{\dfrac{T_1\text{RMSE}_1^2+T_2\text{RMSE}_2^2}{T}}.$$
We also calculate the RMSE across two dimensions: RMSE by age ($\text{RMSE}_i$), which measures the error across age groups for each country and assesses how well the model captures age-specific mortality patterns, and RMSE by year ($\text{RMSE}_t$), which measures the error across years for each country and evaluates the model's ability to capture temporal trends in mortality data. The definitions of $\text{RMSE}_i$ and $\text{RMSE}_t$ are as follows:
$$\text{RMSE}_i^{(1)}=\sqrt{\dfrac{1}{T_1}\sum_{t=1}^{T_1}(m_{ti}^{(1)}-\hat{m}_{ti}^{(1)})^2}, \quad \text{RMSE}_t^{(1)}=\sqrt{\dfrac{1}{N}\sum_{i=1}^N(m_{ti}^{(1)}-\hat{m}_{ti}^{(1)})^2},$$
$$\text{RMSE}_i^{(2)}=\sqrt{\dfrac{1}{T}\sum_{t=1}^T(m_{ti}^{(2)}-\hat{m}_{ti}^{(2)})^2}, \quad \text{RMSE}_t^{(2)}=\sqrt{\dfrac{1}{N}\sum_{i=1}^N(m_{ti}^{(2)}-\hat{m}_{ti}^{(2)})^2},$$
where $m_{ti}^{(1)}=\exp(y_{ti}^{(1)}+a_i^{(1)}), m_{ti}^{(2)}=\exp(y_{ti}^{(2)}+a_i^{(2)})$ represent the mortality rate for age $i$ in year $t$, $\hat{m}_{ti}^{(1)}=\exp(\hat{y}_{ti}^{(1)}+\hat{a}_i^{(1)})$ and $\hat{m}_{ti}^{(2)}=\exp(\hat{y}_{ti}^{(2}+\hat{a}_{i}^{(2)})$ represent the predicted mortality rates across different genders. 

The \textit{fairness} measure is defined as the absolute difference of RMSE between two genders
$$\text{Fairness}=\text{Difference}=|\text{RMSE}_1-\text{RMSE}_2|.$$

\begin{figure}[!h]
    \centering
    \includegraphics[scale=0.5]{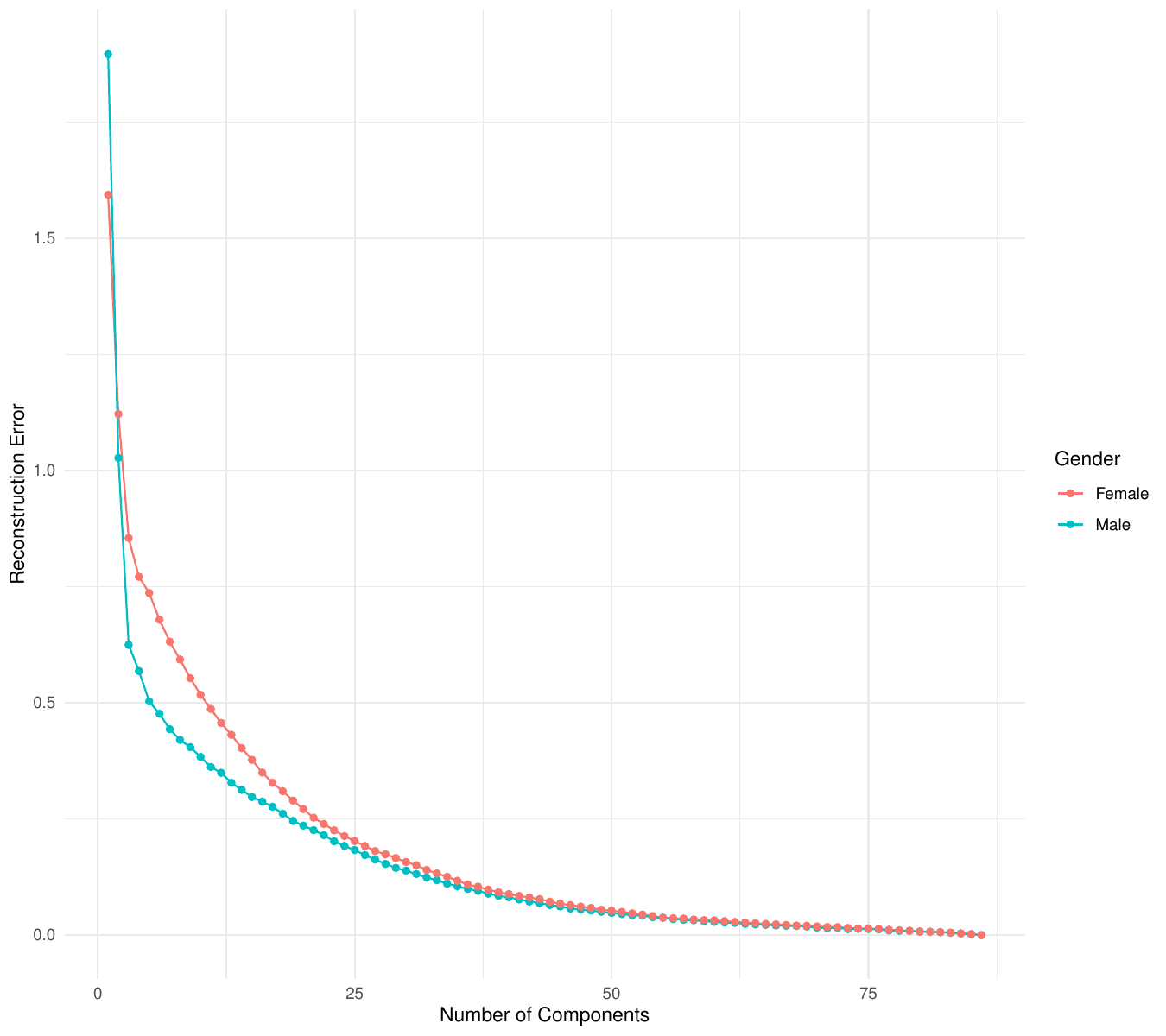}
    \caption{Average reconstruction error of PCA under different values of the number of factors $r$.}
    \label{AURE}
\end{figure}

The reconstruction errors of the factor model in the training set under different numbers of factors $r$ are shown in Figure \ref{AURE}. From this plot, we observe that with a small number of factors, reconstruction errors differ significantly between males and females, with males showing higher reconstruction errors than females. As the number of factors increases, reconstruction errors for both genders gradually converge. Starting from the second component, males show lower reconstruction errors, indicating better reconstruction quality under the same dimensionality. This difference suggests that reconstruction quality differs between genders when using factor models for dimensionality reduction, with males potentially achieving better retention. As factor models are utilized in the dimensionality reduction process, significant disparities in reconstruction error between genders have become apparent. This indicates that fair techniques should be incorporated to improve fairness. In this paper, we select one factor ($r=1$) for empirical analysis, which is also consistent with the \cite{lee-carter}.

\begin{figure}[!h]
    \centering
    \begin{subfigure}{0.9\linewidth}
        \centering
        \includegraphics[width=\textwidth]{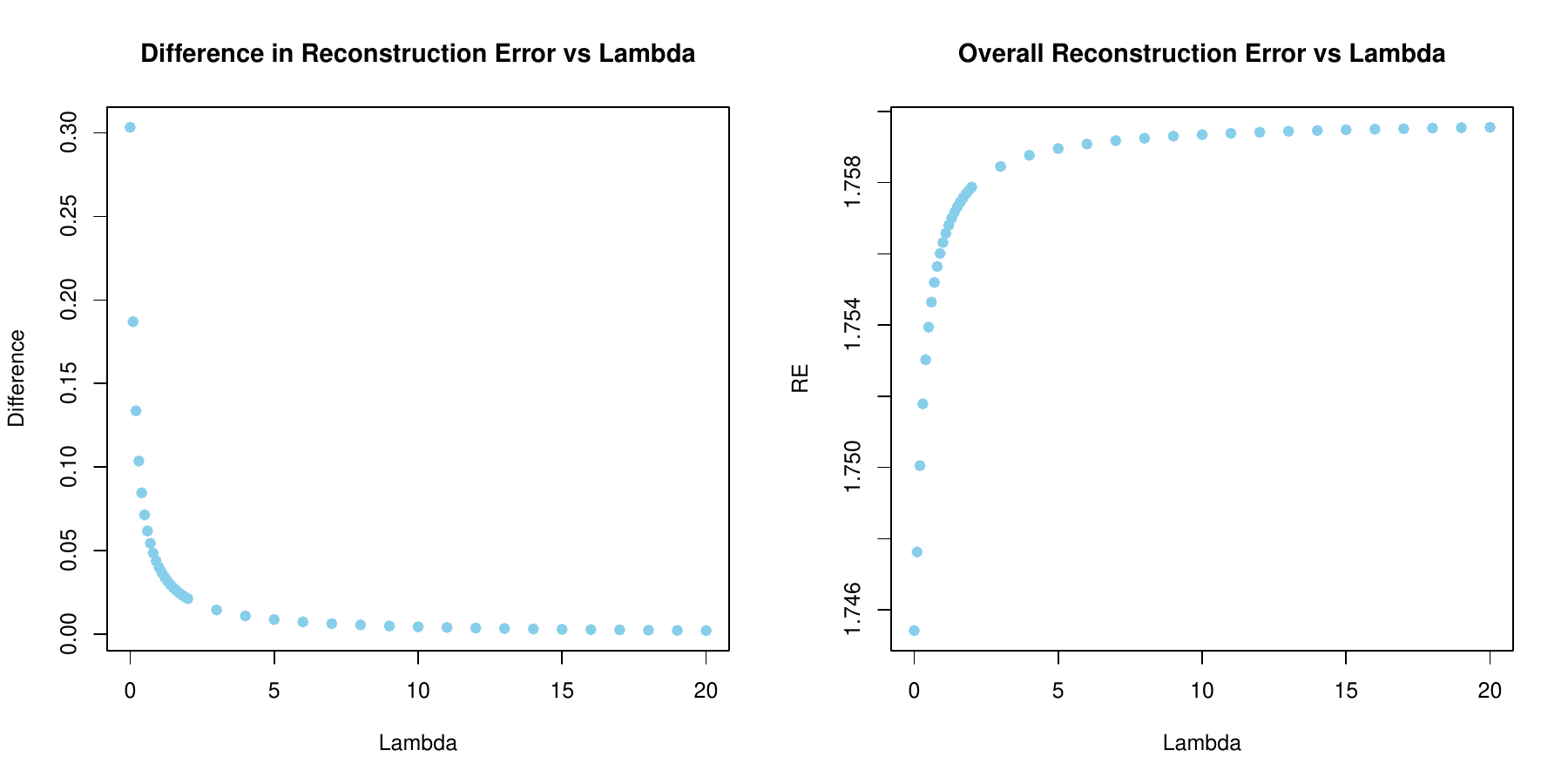}
        \caption{Absolute differences (fairness) between genders and reconstruction error (accuracy) across different $\lambda$ values.}
        \label{RE_Australia}
    \end{subfigure}
    \vspace{0.5cm}     
    \begin{subfigure}{0.9\linewidth}
        \centering
        \includegraphics[width=\textwidth]{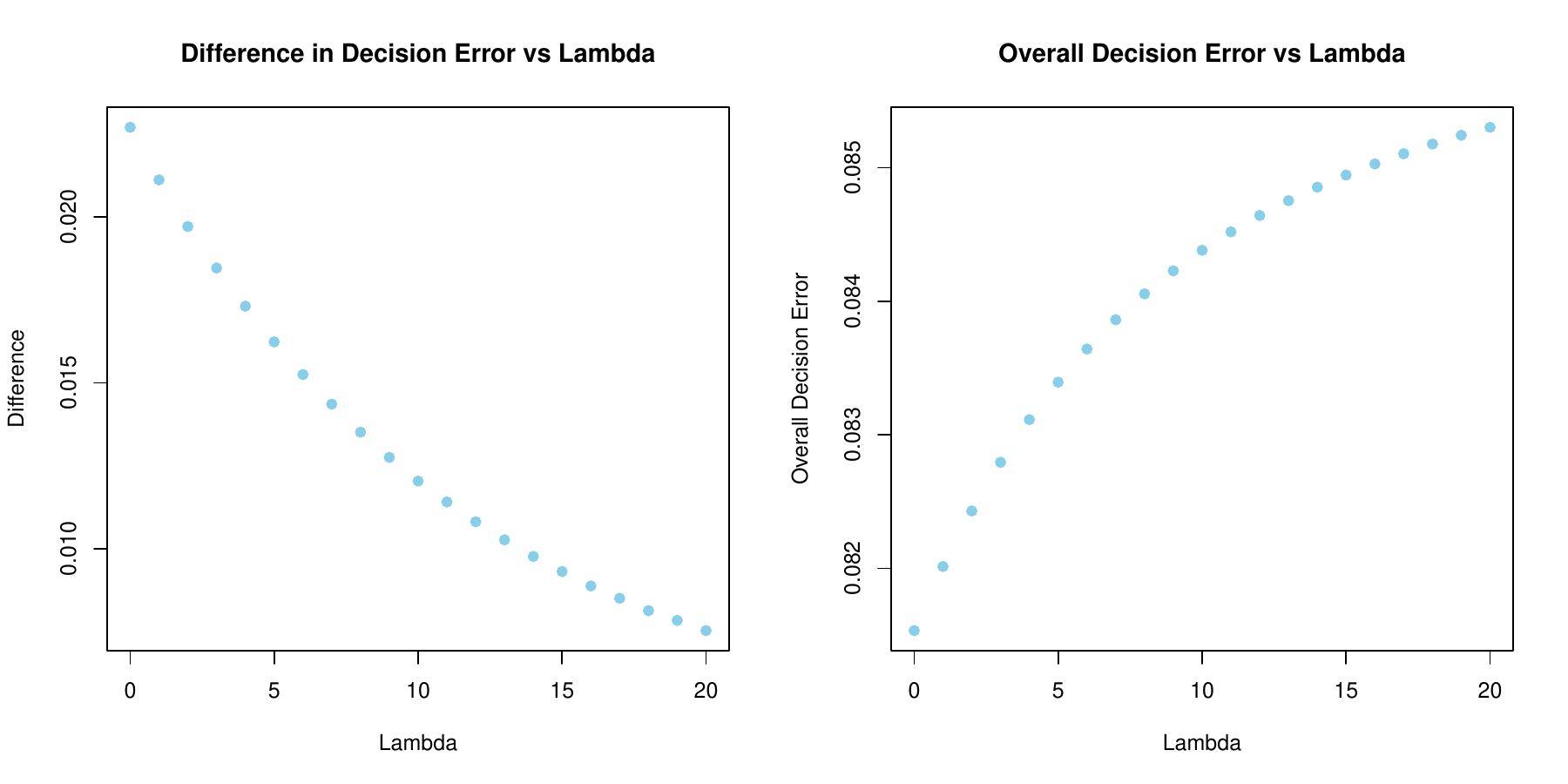}
        \caption{Absolute differences (fairness) between genders and decision error (accuracy) across different $\lambda$ values.}
        \label{DDE_Australia}
    \end{subfigure}
    \caption{Comparison of fairness (absolute differences between genders) and accuracy (errors) in Australia under different $\lambda$ values for the training set.}
    \label{AULambda}
\end{figure}

We first select an appropriate value for $\lambda$ as the penalty parameter for the Fair Factor Model and the Fair Decision Model. Figure \ref{AULambda} illustrates the trade-offs in fairness and accuracy as $\lambda$ varies in the training set. Figure \ref{RE_Australia} shows that for Fair Factor Model with $\lambda = 0$, the difference between males and females is approximately 0.3. Figure \ref{DDE_Australia} illustrates that when $\lambda=0$, the difference in decision errors between males and females is approximately 0.025. As $\lambda$ increases, the difference between the two groups decreases, but the overall error of the data increases.
To finalize the choice of $\lambda$, we performed cross-validation (CV) to evaluate the model’s performance across various values within this range. Ultimately, we selected $\lambda = 11$ as the optimal penalty parameter, balancing regularization strength and predictive accuracy for the Fair Factor Model. $\lambda=2$ is selected for the Fair Decision Model.

\begin{figure}[!h]
    \centering
    \includegraphics[scale=0.5]{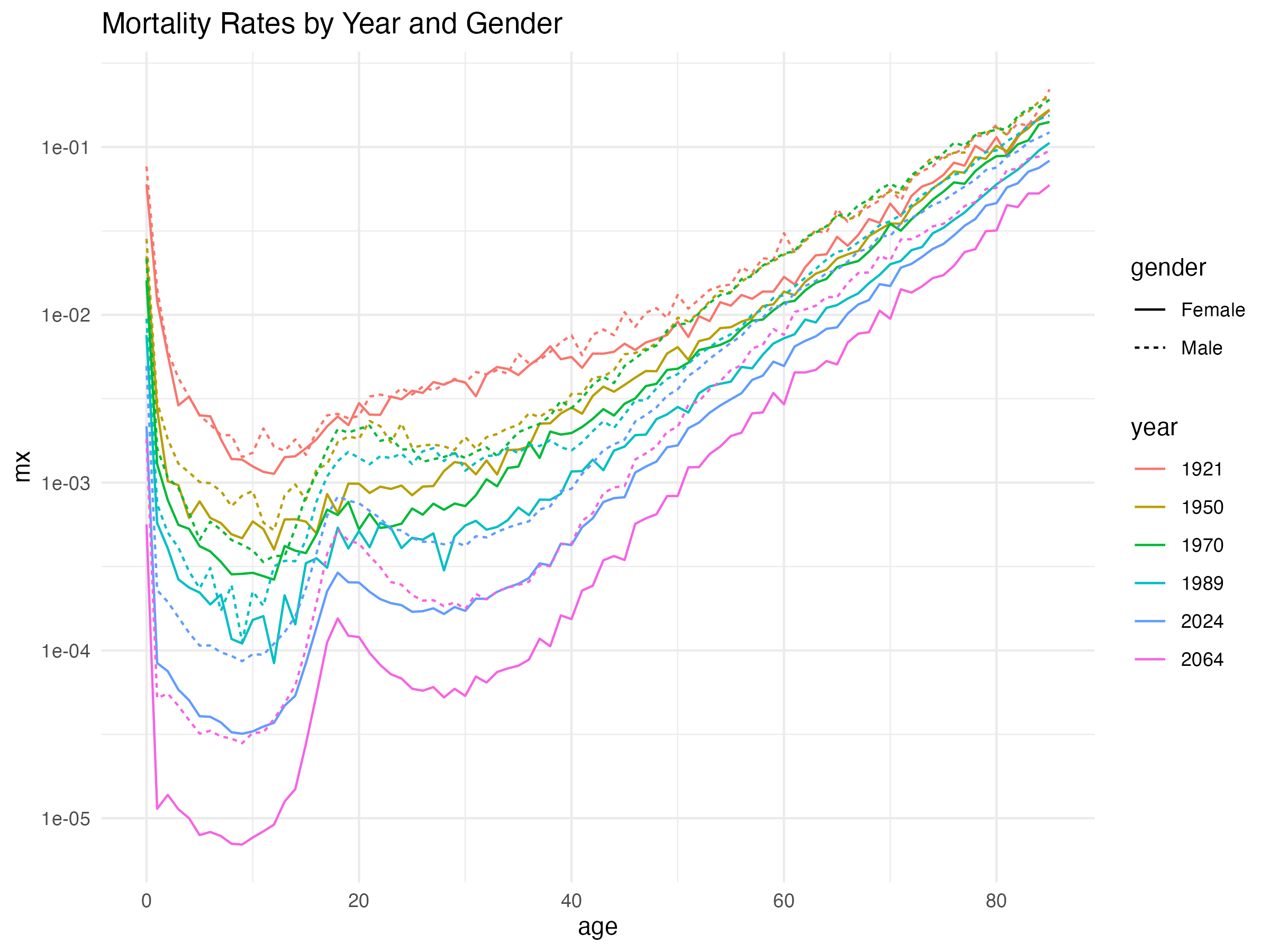}
    \caption{Mortality rate predictions for males and females in Australia using the Fair Factor Model across different ages and years.}
    \label{AUmortality}
\end{figure}

Figure \ref{AUmortality} illustrates the predictive performance of the Fair Factor Model, comparing the estimated mortality rates for both genders at various age groups and time points. From this, we observe that the mortality rate consistently follows a U-shaped trend, with higher rates at the younger and older age groups, lower rates during adolescence and prime age, and a minimum around age 10. Additionally, women consistently exhibit lower mortality rates, consistent with existing research on gender-based differences in mortality rates \citep{10.1373/clinchem.2018.288332}. 
Furthermore, an overall trend of decreasing mortality rates for both men and women over time is observed, likely due to advancements in science and medicine, and this decline is predictable. 

\begin{figure}[!h]
	\centering
	\includegraphics[width=1\linewidth]{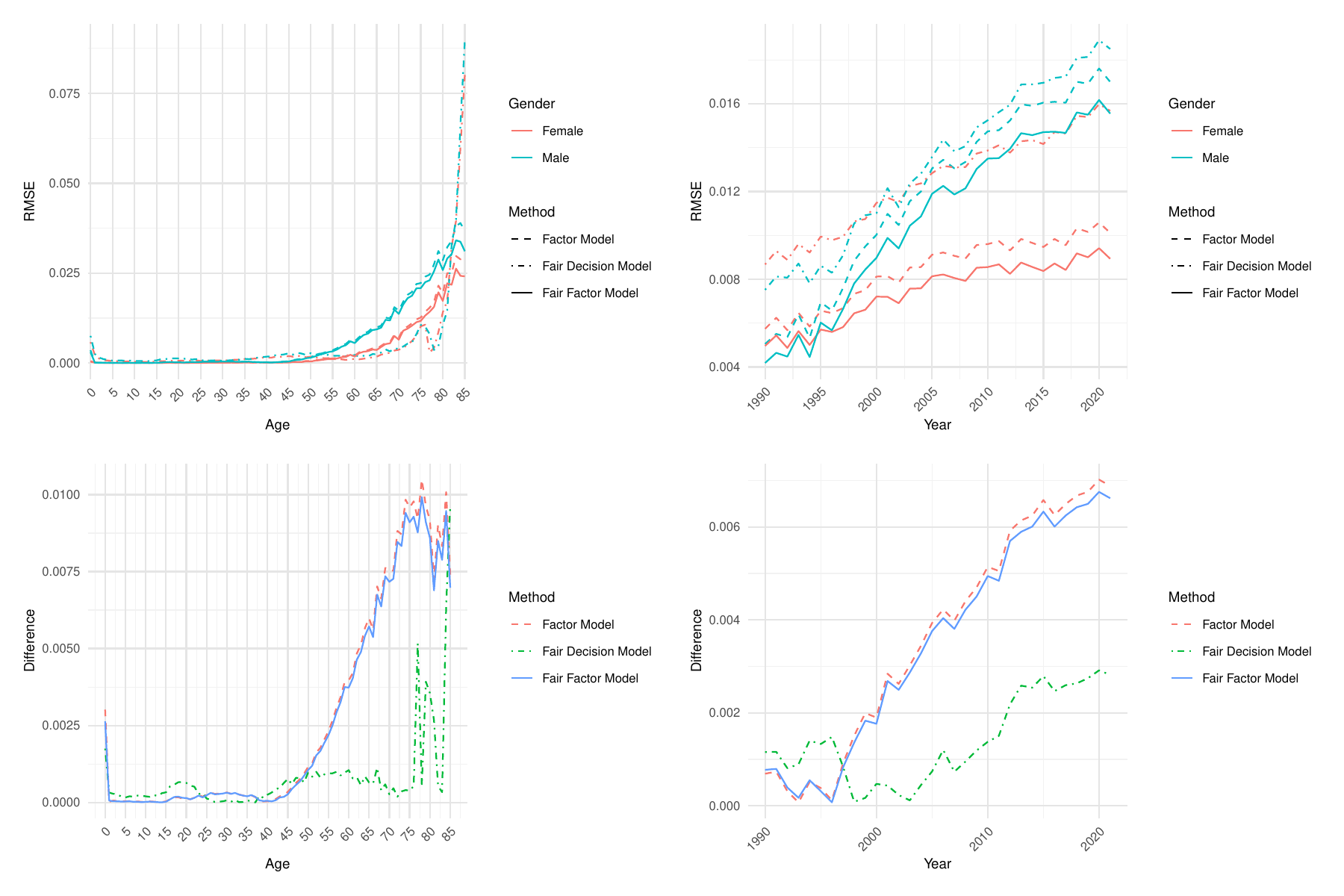}
    \caption{Accuracy and Fairness of Mortality Predictions Across Age and Time Using Different Models.
Top panels: Prediction accuracy (RMSE of mortality rates) by age and year for the Factor Model, Fair Factor Model, and Fair Decision Model.
Bottom panels: Fairness in predictions, measured by the absolute RMSE difference between genders, across age and year, illustrating the extent of gender disparities.}
	\label{AUprediction}
\end{figure}

Figure \ref{AUprediction} presents the accuracy and fairness of mortality rate predictions for males and females across different ages and years using Factor Model, Fair Factor Model and Fair Decision Model. The left column of Figure \ref{AUprediction} shows age-specific comparison plots. We see that although the Fair Factor Model improves prediction performance compared to the Factor Model, the fairness between males and females remains limited after incorporating $\lambda$, especially beyond age 45, where the gender difference remains noticeable. The Fair Decision Model introduces a higher degree of fairness and yields a lower RMSE than both the Factor Model and Fair Factor Model. Although the small sample size in the highest age groups may produce higher prediction errors, the Fair Decision Model still demonstrates strong performance in both fairness and accuracy overall.

The year-specific plots are shown in the right column of Figure \ref{AUprediction}. We see that although the Fair Factor Model effectively reduces RMSE over the long term compared to the Factor Model, it does not visibly introduce fairness in this plot. During these years, the Fair Decision Model sacrifices prediction accuracy; however, it significantly enhances fairness between males and females in mortality. 

The graph in the lower-left corner of Figure \ref{AUprediction} compares the fairness (i.e. 
 absolute RMSE difference) between males and females for the Factor Model, Fair Factor Model, and Fair Decision Model. Comparing the results reveals that the lines for the Factor Model and Fair Factor Model overlap significantly, indicating that the Fair Factor Model does not introduce significant fairness between males and females. The Fair Decision Model shows a marked effect in reducing the prediction error difference between males and females, indicating an effective improvement in fairness. The lower-right panel demonstrates a similar pattern, with a large gap between the Fair Decision Model and the other models after 2000, indicating that the Fair Decision Model further enhances fairness in predictions.

\begin{table}[!h]
	\caption{Overall accuracy (RMSE) and overall fairness (Difference) in mortality by gender.}\label{mortality_AUS}
	\renewcommand{\arraystretch}{1.2}
	\centering
	\scalebox{1}{
		\begin{tabular}{ccccccccc}
			\toprule[2pt]
			RMSE&Factor Model &Fair Factor Model &Fair Decision Model\\
			  \cmidrule(r){1-4}   
			Accuracy (Male) &0.0126 &0.0115 &0.0137 \\
                Accuracy (Female) &0.0086 &0.0076 &0.0126 \\
                \cmidrule(r){1-4}
                Fairness (Difference) &0.0041 &0.0039 &0.0011 \\
                \cmidrule(r){1-4}
                Total Accuracy &0.0108 &0.0097 &0.0132 \\
			\bottomrule[2pt]
		\end{tabular}
        }
\end{table}

Table \ref{mortality_AUS} presents the predictive accuracy (RMSE) and fairness performance by gender for various mortality modeling techniques. We can see that the Fair Decision Model sacrifices female prediction accuracy to improve fairness. It successfully increases fairness between males and females (0.0011), compared to the other techniques (0.0041 for Factor Model and 0.0039 for Fair Factor Model). However, the accuracy (RMSE) of the Fair Decision Model is 0.0132, which exceeds that of the Factor Model (0.0108) and the Fair Factor Model (0.0097). This discrepancy arises because the Fair Decision Model primarily aims to improve decisions (i.e. annuity values) rather than mortality predictions. 

In conclusion, the Fair Factor Model demonstrates the best performance in terms of accuracy and the second-best in terms of fairness. Compared to the standard Factor Model, it achieves notable improvements in both prediction accuracy and fairness. Notably, accuracy and fairness are achieved simultaneously: the fairness regularization not only promotes fairness but also reduces model complexity, leading to a better bias–variance trade-off. In contrast, the Fair Decision Model sacrifices a substantial amount of accuracy in pursuit of fairness. Nevertheless, its performance in mortality prediction remains reasonable, as its primary objective is to reduce discrimination in annuity valuation rather than to optimize mortality estimates directly.

\subsection{Fair Annuity Pricing}
In this section, the expected present value (EPV) of an annuity-due is generated using the three methods described in Section \ref{Mortality Application}.   For both the Factor Model and the Fair Factor Model, mortality rates are first predicted using their respective methods, and the EPV of an annuity-due is then calculated based on these predictions. In contrast, the Fair Decision Model applies fairness constraints directly to the EPV of the annuity-due. The predicted present values are compared to the actual EPVs computed from observed mortality rates, enabling an evaluation of prediction accuracy across genders and an assessment of each model’s effectiveness in promoting fairness.


Based on the mortality rates predicted using different methods from 1990 to 2021 for different ages, we calculated the EPV of the annuity-due for a term of $n=10$ and compared it with the EPV derived from the actual mortality rates. We compare the accuracy and fairness of different methods using similar measures discussed in Section \ref{Mortality Application}.  By replacing mortality rates, $m_{ti}$, with the expected present value of an annuity-due, $p_{ti}(\bbm_t)$, we derive the accuracy and fairness metrics for annuity pricing. 

\begin{figure}[!h]
	\centering
	\includegraphics[width=1\textwidth]{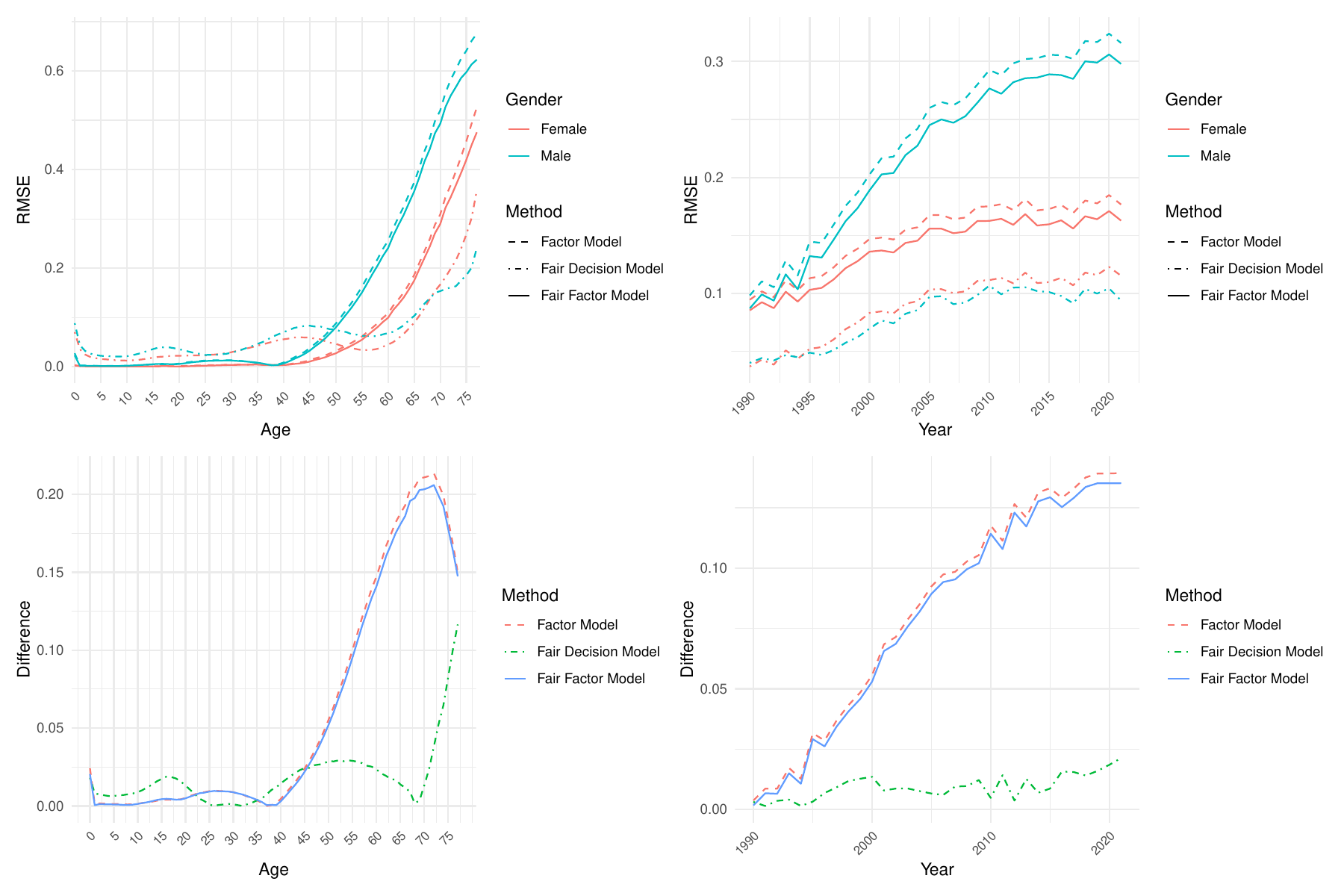}
	\caption{Accuracy and Fairness of Annuity Pricing Using Different Modeling Approaches.
Top panels: Prediction accuracy of annuity prices (RMSE of the expected present value of an annuity-due) by age and  year  for the Factor Model, Fair Factor Model, and Fair Decision Model.
Bottom panels: Fairness in annuity pricing, measured by the absolute RMSE difference between genders, by age and year, comparing the three methods.}
	\label{RMSE_annuity_Au}
\end{figure}

Figure \ref{RMSE_annuity_Au} presents a comparison of both accuracy and fairness in the EPV of an annuity-due, as generated by different models across genders, ages, and years. Compared to the standard Factor Model, the Fair Factor Model produces more accurate and fairer results over both age groups and time periods. Among the three models, the Fair Decision Model achieves the best overall performance in terms of both accuracy and fairness.

The left column of Figure \ref{RMSE_annuity_Au} presents age-specific comparison plots. Before age 50, the Fair Decision Model exhibits slightly higher RMSE than the other models. However, after age 50, it provides more accurate EPV predictions—particularly for males. The fairness plot (based on the absolute difference in RMSE between genders) shows that the Fair Decision Model achieves substantially greater fairness compared to the other models. In the right column of Figure \ref{RMSE_annuity_Au}, which displays year-specific comparisons, the Fair Decision Model consistently delivers the most accurate and fair results across the entire time period.

The superior performance of the Fair Decision Model is further demonstrated in Table \ref{annuity_AUS}. As shown in the table, prediction accuracy improves for both males and females with the introduction of the Fair Decision Model. The total RMSE is 0.0882, which is lower than that of the other models, indicating accurate predictions of the EPV of an annuity-due. In terms of fairness, the total difference in RMSE between genders is 0.0095 for the Fair Decision Model, compared to 0.0916 for the Factor Model and 0.0886 for the Fair Factor Model. This highlights the strong performance of the Fair Decision Model in achieving both accuracy and fairness. Similar to the discussion in Section \ref{Mortality Application}, this dual improvement arises because the fairness regularization not only promotes equity but also reduces model complexity, thereby improving the bias–variance trade-off.

\begin{table}[!h]
	\caption{Overall accuracy (RMSE) and overall fairness (Difference) in annuity by gender}\label{annuity_AUS}
	\renewcommand{\arraystretch}{1.2}
	\centering
	\scalebox{1}{
		\begin{tabular}{ccccccccc}
			\toprule[2pt]
			RMSE&Factor Model &Fair Factor Model &Fair Decision Model\\
			  \cmidrule(r){1-4}   
			Accuracy (Male) &0.2453 &0.2306 &0.0833\\
                Accuracy (Female) &0.1537 &0.1420 &0.0928\\
                \cmidrule(r){1-4}
                Fairness (Difference) &0.0916 &0.0886 &0.0095\\
                \cmidrule(r){1-4}
                Total Accuracy &0.2047 &0.1915 &0.0882 \\
			\bottomrule[2pt]
		\end{tabular}
        }
\end{table}

\begin{figure}[!h]
	\centering
	\includegraphics[width=1\textwidth]{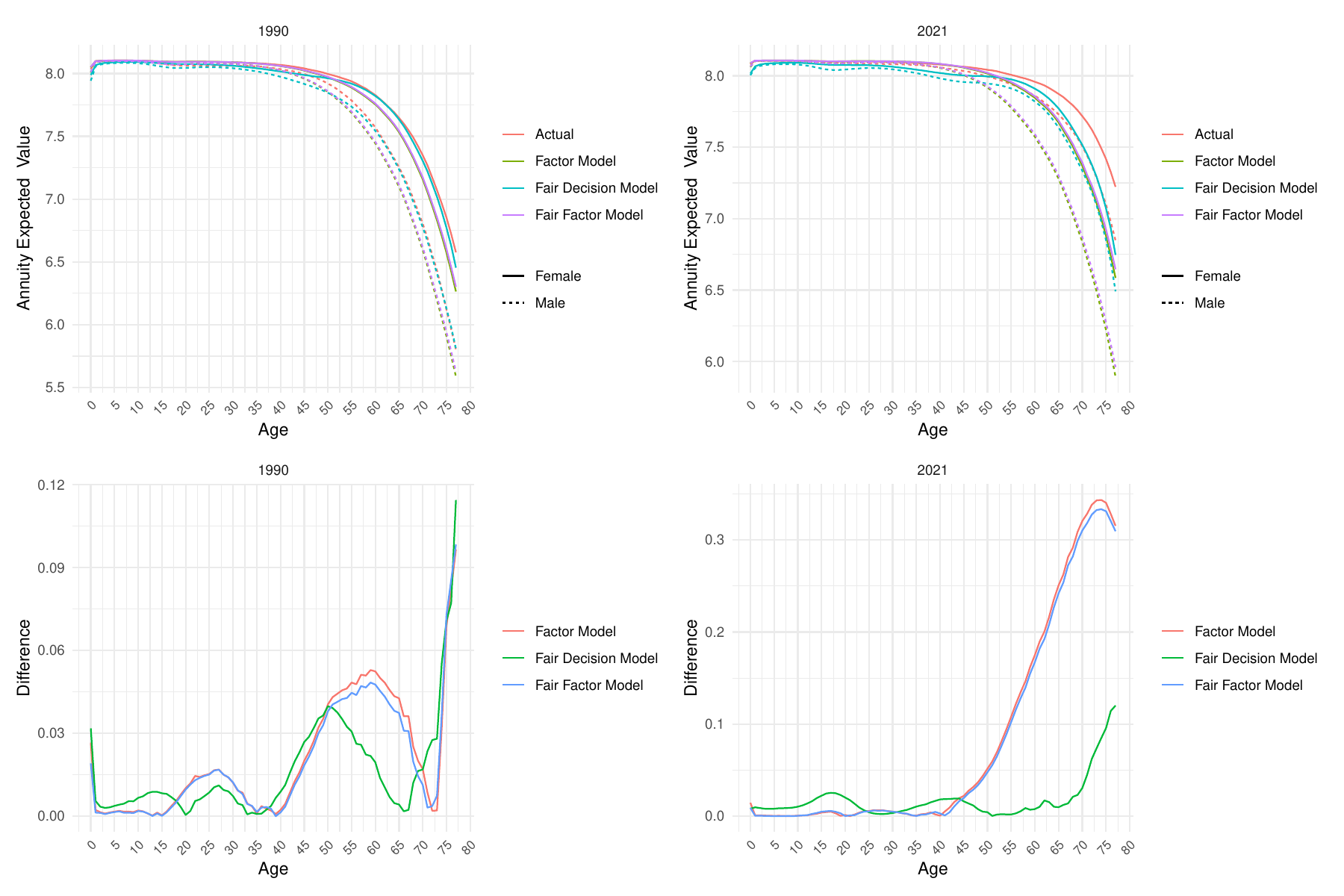}
	\caption{Comparison of Annuity-Due Valuation Across Methods.
Top panels: EPVs of an annuity-due estimated by the Factor Model, Fair Factor Model, and Fair Decision Model, compared to EPVs computed using actual mortality rates, for the years 1990 and 2021 ($n = 10$).
Bottom panels (Fairness): Absolute differences in EPV estimation errors between genders, highlighting fairness discrepancies across methods in 1990 and 2021.}
	\label{AUannuity}
\end{figure}

Figure \ref{AUannuity} presents a comparison of the present value of annuities estimated by the three methods for both genders, alongside the actual present values derived from observed mortality rates in 1990 and 2021. The figure also includes the absolute differences between the estimated and actual EPVs, allowing for an assessment of both accuracy and fairness across genders.

In the top panels, which display the EPV of an annuity-due, all models tend to underestimate the actual present values, particularly for older age groups. For the 1990 cohort, the Factor Model and Fair Factor Model yield similar predictions for both males and females. In contrast, the Fair Decision Model provides estimates that are closer to the actual EPVs for both genders. In 2021—the final year of the time series—all models show decreased prediction accuracy, which is expected due to the nature of time-series forecasting. Nevertheless, the Fair Decision Model continues to outperform the others in terms of predictive accuracy.

The bottom panels of Figure \ref{AUannuity} display fairness plots, defined as the absolute difference in RMSE between genders. In 1990, all three models exhibit similar trends across age groups. However, the Fair Decision Model achieves lower gender-based differences in specific age ranges, notably between ages 20–35 and 50–70, suggesting better fairness. In 2021, while the Factor Model and Fair Factor Model follow a similar pattern, the Fair Decision Model consistently maintains a lower difference between male and female predictions. This indicates that the Fair Decision Model is more effective at promoting fairness in long-term annuity price predictions.

Using the standard Factor Model, for an annuity-due with annual payments of \$100,000 over 10 years, the average absolute difference in the RMSE of annuity prices between genders ranges from approximately \$10,000 to \$40,000 for individuals aged 55 to 80 in 2021, highlighting a substantial discrepancy in the accuracy of annuity pricing across genders. However, this disparity is significantly reduced when applying the Fair Decision Model, with differences falling below \$15,000, demonstrating a notable improvement in fairness and a substantial reduction in gender-based discrepancies in pricing accuracy.

In summary, applying the three models to Australian mortality data to generate the EPV of an annuity-due for males and females reveals clear performance differences. The Fair Decision Model demonstrates strong performance in improving both accuracy and fairness across genders, particularly in long-term time series predictions. While the Fair Factor Model achieves the highest accuracy in mortality prediction, its accuracy and fairness are not as strong as that of the Fair Decision Model for annuity pricing.

\section{Conclusion}
This paper addresses the challenge of fairness in mortality modeling and annuity pricing when gender—a protected but often legally permitted attribute—is explicitly used in actuarial models. We demonstrate that standard factor models, such as the Lee-Carter model, can exhibit systematic discrepancies in prediction accuracy across genders, which in turn lead to disparities in annuity pricing outcomes.

To mitigate these disparities, we introduce two fairness-aware extensions of the traditional factor model. The Fair Factor Model incorporates fairness regularization into the estimation process, achieving a better balance in prediction errors across demographic groups while also potentially improving overall accuracy. The Fair Decision Model goes a step further by directly targeting fairness in downstream decisions, such as annuity valuation, through a novel criterion we call Decision Error Parity.

Our empirical analysis using Australian mortality data confirms that the Fair Decision Model achieves the best overall performance in terms of both fairness and pricing accuracy for annuities. While it sacrifices some predictive accuracy in mortality rates, it excels in reducing gender-based disparities in pricing outcomes—highlighting the importance of aligning fairness objectives with real-world decision-making contexts.

These findings emphasize that fairness should be addressed not only at the prediction stage, but also at the level of decision-making where model outputs are applied. Our proposed models provide a principled and practical framework for incorporating fairness into actuarial modeling workflows, offering new tools for regulators and practitioners committed to equitable outcomes in insurance and pension systems.

Future work may extend this framework to multi-class protected attributes, incorporate causal fairness constraints, or explore applications in broader actuarial and financial domains.

\bibliographystyle{agsm}
\bibliography{main}

\end{document}